\documentstyle[epsfig]{mn}

\newif\ifAMStwofonts
\newcommand{\target}{XTE\,J1118+480}
\newcommand{\HST} {\textit{HST}}
\newcommand{\XTE} {\textit{RXTE}}
\newcommand{\RXTE}{\textit{RXTE}}
\newcommand{\EUVE}{\textit{EUVE}}
\newcommand{\Chandra}{\textit{Chandra}}

\newcommand{\cntpersec}{\,cnt\,s$^{-1}$}
\ifoldfss
  \ifCUPmtlplainloaded \else
    \NewTextAlphabet{textbfit} {cmbxti10} {}
    \NewTextAlphabet{textbfss} {cmssbx10} {}
    \NewMathAlphabet{mathbfit} {cmbxti10} {} % for math mode
    \NewMathAlphabet{mathbfss} {cmssbx10} {} %  "   "    "
  \fi
  \ifAMStwofonts
    \ifCUPmtlplainloaded \else
      \NewSymbolFont{upmath} {eurm10}
      \NewSymbolFont{AMSa} {msam10}
      \NewMathSymbol{\upi}     {0}{upmath}{19}
      \NewMathSymbol{\umu}     {0}{upmath}{16}
      \NewMathSymbol{\upartial}{0}{upmath}{40}
      \NewMathSymbol{\leqslant}{3}{AMSa}{36}
      \NewMathSymbol{\geqslant}{3}{AMSa}{3E}

      \let\geq=\geqslant 
    \fi
  \fi
\fi % End of OFSS

\ifnfssone
  \newmathalphabet{\mathit}
  \addtoversion{normal}{\mathit}{cmr}{m}{it}
  \addtoversion{bold}{\mathit}{cmr}{bx}{it}
  \newmathalphabet{\mathbfit} % math mode version of \textbfit{..}
  \addtoversion{normal}{\mathbfit}{cmr}{bx}{it}
  \addtoversion{bold}{\mathbfit}{cmr}{bx}{it}
  \newmathalphabet{\mathbfss} % math mode version of \textbfss{..}
  \addtoversion{normal}{\mathbfss}{cmss}{bx}{n}
  \addtoversion{bold}{\mathbfss}{cmss}{bx}{n}
  \ifAMStwofonts
    \ifCUPmtlplainloaded \else
      \UseAMStwoboldmath
      \makeatletter
      \new@mathgroup\upmath@group
      \define@mathgroup\mv@normal\upmath@group{eur}{m}{n}
      \define@mathgroup\mv@bold\upmath@group{eur}{b}{n}
      \edef\UPM{\hexnumber\upmath@group}
      \new@mathgroup\amsa@group
      \define@mathgroup\mv@normal\amsa@group{msa}{m}{n}
      \define@mathgroup\mv@bold\amsa@group{msa}{m}{n}
      \edef\AMSa{\hexnumber\amsa@group}
      \makeatother
      \mathchardef\upi="0\UPM19
      \mathchardef\umu="0\UPM16
      \mathchardef\upartial="0\UPM40
      \mathchardef\leqslant="3\AMSa36
      \mathchardef\geqslant="3\AMSa3E

      \let\geq=\geqslant 
    \fi
  \fi
\fi % End of NFSS release 1

\ifnfsstwo
  \DeclareMathAlphabet{\mathbfit}{OT1}{cmr}{bx}{it}
  \SetMathAlphabet\mathbfit{bold}{OT1}{cmr}{bx}{it}
  \DeclareMathAlphabet{\mathbfss}{OT1}{cmss}{bx}{n}
  \SetMathAlphabet\mathbfss{bold}{OT1}{cmss}{bx}{n}
  \ifAMStwofonts
    \ifCUPmtlplainloaded \else
      \DeclareSymbolFont{UPM}{U}{eur}{m}{n}
      \SetSymbolFont{UPM}{bold}{U}{eur}{b}{n}
      \DeclareSymbolFont{AMSa}{U}{msa}{m}{n}
      \DeclareMathSymbol{\upi}{0}{UPM}{"19}
      \DeclareMathSymbol{\umu}{0}{UPM}{"16}
      \DeclareMathSymbol{\upartial}{0}{UPM}{"40}
      \DeclareMathSymbol{\leqslant}{3}{AMSa}{"36}
      \DeclareMathSymbol{\geqslant}{3}{AMSa}{"3E}

      \let\geq=\geqslant 
    \fi
  \fi
\fi % End of NFSS release 2

\ifCUPmtlplainloaded \else
  \ifAMStwofonts \else % If no AMS fonts
    \def\upi{\pi}
    \def\umu{\mu}
    \def\upartial{\partial}
  \fi
\fi

\title[Rapid multi-wavelength variability in XTE J1118+480]
       {The remarkable rapid X-ray, ultraviolet\thanks{Based on
observations made with the NASA/ESA Hubble Space Telescope, obtained
at the Space Telescope Science Institute, which is
operated by the Association of Universities for Research in Astronomy, 
Inc., under NASA contract NAS 5-26555. These observations are associated with
proposal GO\,8647.}, optical, and IR variability 
	in the black hole XTE\,J1118+480}
\author[R. I. Hynes et al.]
       {R. I. Hynes$^{1,2}$\thanks{E-mail: rih@astro.as.utexas.edu;
       Hubble Fellow},
	C. A. Haswell$^3$, 
	W. Cui$^{4}$, 
	C. R. Shrader$^5$,
	K. O'Brien$^{6,7,8}$, \newauthor
	S. Chaty$^{3,9,10}$,
	D. R. Skillman$^{11}$, 
        J. Patterson$^{12}$, 
	Keith Horne$^{7}$\\
	$^1$Astronomy Department, The University of Texas at Austin, 1
        University Station C1400, Austin, Texas 78712-0259, USA\\
        $^2$Department of Physics and Astronomy, University of
	Southampton, Southampton, SO17 1BJ, UK\\
	$^3$Department of Physics and Astronomy, The Open University,
        Walton Hall, Milton Keynes, MK7 6AA, UK\\
        $^4$Department of Physics, Purdue University, 1396 Physics Building,
        West Lafayette, IN 47907-1396, USA\\
	$^5$Laboratory for High-Energy Astrophysics, NASA
        Goddard Space Flight Center, Greenbelt, MD 20771, USA\\
	$^6$Sterrenkundig Instituut, Kruislaan 403, 1098 SJ Amsterdam,
	The Netherlands\\ 
	$^7$School of Physics and Astronomy, University of St. Andrews, 
    	North Haugh, St. Andrews, Fife KY16 9SS, UK\\
	$^8$European Southern Observatory, Casilla 19001,
	Santiago 19, Chile\\
$^9$ {Universit\'e Paris 7, F\'ed\'eration APC, 2 place Jussieu, 75005 
Paris, France} \\
$^{10}$ {Service d'Astrophysique,
DSM/DAPNIA/SAp, CEA-Saclay, Bat. 709, L'Orme des Merisiers
F-91 191 Gif-sur-Yvette, Cedex, France} \\
	$^{11}$Center for Backyard Astrophysics (East), 9517
        Washington Avenue, Laurel, MD 20723, USA\\
	$^{12}$Department of Astronomy, Columbia University, 
        550 West 120th Street, New York, NY 10027, USA}
\date{Accepted ??.
      Received ??;
      in original form ??}

\pagerange{\pageref{firstpage}--\pageref{lastpage}}
\pubyear{2003}

\begin{document}

\maketitle

\label{firstpage}

\begin{abstract}
The transient black hole binary \target\ exhibited dramatic rapid
variability at all wavelengths which were suitably observed during its
2000 April--July outburst.  We examine time-resolved X-ray,
ultraviolet, optical, and infrared data spanning the plateau phase of
the outburst.  We find that both X-ray and infrared bands show large
amplitude variability.  The ultraviolet and optical variability is
more subdued, but clearly correlated with that seen in the X-rays.
The ultraviolet, at least, appears to be dominated by the continuum,
although the lines are also variable.  Using the X-ray variations as a
reference point, we find that the UV variability at long wavelengths
occurs later than that at short wavelengths.  Uncertainty in \HST\
timing prohibits a determination of the absolute lag with respect to
the X-rays, however.  The transfer function is clearly not a
$\delta$-function, exhibiting significant repeatable structure.  For
the main signal we can rule out an origin in reprocessing on the
companion star -- the lack of variation in the lags is not consistent
with this given a relatively high orbital inclination.  Weak
reprocessing from the disc and/or companion star may be present, but
is not required, and another component must dominate the variability.
This could be variable synchrotron emission correlated with X-ray
variability, consistent with our earlier interpretation of the IR flux
as due to synchrotron emission rather than thermal disc emission.  In
fact the broad-band energy distribution of the variability from IR to
X-rays is consistent with expectations of optically thin synchrotron
emission.  We also follow the evolution of the low-frequency
quasi-periodic oscillation in X-rays, UV, and optical.  Its properties
at all wavelengths are similar indicating a common origin.
\end{abstract}

\begin{keywords}
accretion, accretion discs -- binaries: close -- stars: individual: 
XTE J1118+480 -- ultraviolet: stars -- X-rays: stars
\end{keywords}
%
%%%%%%%%%%%%%%%%%%%%%%%%%%%%%%%%%%%%%%%%%%%%%%%%%%%%%%%%%%%%%%%%%%%%%%%%%%%%%%%
%
\section{Introduction}
Black hole X-ray transients (BHXRTs), also referred to as X-ray novae
and soft X-ray transients (Tanaka \& Shibazaki 1996; Cherepashchuk
2000), are a class of low-mass X-ray binaries (LMXBs) in which long
periods of quiescence, typically decades, are punctuated by very
dramatic X-ray and optical outbursts, frequently accompanied by radio
activity.  Often the X-ray emission is dominated by thermal emission
from the hot inner accretion disc, and UV/optical emission is thought
to be produced by reprocessing of X-rays.  This means that some or all
of the UV/optical variability is actually reprocessed X-ray
variability.

Since there is a finite light travel time from the X-ray source to the
reprocessing region, a finite lag is expected between X-rays and
UV/optical emission.  By measuring this lag we can hope to measure
spatial scales within the binary.  This is the essence of echo-mapping
in X-ray binaries (O'Brien \& Horne 2001; O'Brien et al.\ 2002).  The
method has great promise; for example it could measure the binary
separation if orbital phase-resolved echoes from the companion star
are seen.  It has so far had very little application to Galactic black
holes.  For the BHXRT GRO~J1655--40 (Hynes et al.\ 1998; O'Brien et
al.\ 2002), a smeared, lagged response (with a delay 10--20\,s) was
found, a delay too short for light travel times to the companion star.
The correlated signal was attributed to reprocessing in the accretion
disc, but the data quality prohibited a detailed reconstruction of the
disc transfer function.  A very short coordinated optical and X-ray
observation of GX~339--4 was also made (Motch et al.\ 1983).  There,
however, reprocessing was ruled out as an origin for the optical
variability.

Another benefit of obtaining high time-resolution multi-wavelength
data is to search for quasi-periodic oscillations (QPOs) in several
wavebands.  To date these have been studied most extensively at X-ray
energies (van der Klis 1995, 2000) where enormous progress has been
recently made thanks to copious high time resolution data from the
{\it Rossi X-ray Timing Explorer} (\RXTE).  Optical studies are rarer
and confined to low frequency QPOs in GX~339--4 (e.g.\ Motch,
Ilovaisky \& Chevalier 1982; Motch et al.\ 1983).

\target\ was discovered by the {\it RXTE} All Sky Monitor (ASM) on
2000 March 29 (Remillard et al.\ 2000) as a weak, slowly rising X-ray
source.  Analysis of earlier data revealed a previous outburst in 2000
January reaching a similar brightness.  A power-law spectrum was seen
out to at least 120\,keV (Wilson \& McCollough 2000), with spectral
index similar to Cyg X-1 in the low/hard state.  \target\ is also a
significant radio source, first detected at 6.2\,mJy (Pooley \&
Waldram 2000).  A 13th magnitude optical counterpart was promptly
identified, where the previously known brightest object was at 18.8
(Uemura, Kato \& Yamaoka 2000; Uemura et al.\ 2000).  The optical
spectrum was typical of BHXRTs in outburst (Garcia et al.\ 2000).
Continued observations revealed a weak photometric modulation on a
4.1\,hr period (Cook et al.\ 2000); shorter than any known orbital
period among black hole candidates.  This was likely due to
superhumps.  Subsequent quiescent observations have confirmed an
orbital period shorter than the outburst period by $<1$\,per cent
(Zurita et al.\ 2002).  The mass function has been measured to be
$6.1\pm0.3$\,M$_{\odot}$ (McClintock et al.\ 2001a; Wagner et al.\
2001), making \target\ a secure black hole candidate.

The optical brightness in outburst was somewhat surprising, as the
X-rays were so faint.  It was initially suggested that the system
might be at very high inclination, so that the X-ray source was
obscured by the disc rim and only scattered X-rays were visible
(Garcia et al.\ 2000).  However no eclipses were seen, and the EUV
flux did not show significant orbital modulation (Hynes et al.\ 2000),
so the inclination, although high (71--82$^{\circ}$, Zurita et al.\
2002), is not high enough for the central source to be obscured by the
disc rim.  Consequently the low X-ray brightness must be because the
source was intrinsically faint in X-rays; it was in a rather low
`low/hard' state (Hynes et al.\ 2000).  This makes the outburst
unusual for a BHXRT.  The X-ray emission contains no detectable disc
component (Hynes et al.\ 2000; McClintock et al.\ 2001b) and the UV,
optical, and IR (UVOIR) probably have significant contributions from
synchrotron emission (e.g.\ Hynes et al.\ 2000; Merloni, Di Matteo \&
Fabian 2000; Fender et al.\ 2001; Chaty et al.\ 2003).  It is
therefore questionable whether echo mapping will be applicable to this
source.  Nonetheless, clear, lagged correlations are seen between
rapid X-ray and UV/optical variations (Haswell et al.\ 2000; Kanbach
et al.\ 2001).  It is of considerable interest to explore whether
these correlations do indeed arise from reprocessing in the disc
and/or companion star or from some other mechanism, as suggested by
Merloni et al.\ (2000) and Kanbach et al.\ (2001).

\target\ lies at very high Galactic latitude ($+62\degr$) and is close
to the Lockman hole (Lockman, Jahoda \& McCammon 1986).  This results
in a very low interstellar absorption ($N_{\rm H} \sim (0.7-1.3)\times
10^{20}$\,cm$^{-2}$, Hynes et al.\ 2000; McClintock et al.\ 2001b;
Chaty et al.\ 2003).  This, together with its brightness, made it an
ideal target for multi-wavelength studies of correlated variability.
This work will focus only upon timing results from our
multi-wavelength campaign.  Other results from the campaign are
described by Hynes et al.\ (2000), Haswell et al.\ (2002) and Chaty et
al.\ (2003); see also McClintock et al.\ (2001b) and Esin et al.\
(2001).
%
%%%%%%%%%%%%%%%%%%%%%%%%%%%%%%%%%%%%%%%%%%%%%%%%%%%%%%%%%%%%%%%%%%%%%%%%%%%%%%%
\section{Observations}
\begin{table*}
\caption{Log of outburst time-resolved observations.  For \HST\ and
\XTE\ data, the rms values are based on 1\,s binned lightcurves.  For
UKIRT they are based on the individual images.  The Source rms column
is defined by $\sigma_{\rm source}^2 = \sigma_{\rm total}^2 -
\sigma_{\rm noise}^2$.  No rms values are quoted for the last \HST\
observations of June 24--25 as this was contaminated by spurious low
frequency variations.  Neither have they been calculated for
individual optical observations; typical values for these data are a
source rms around $\sim20$\,percent and errors much less than this.
For \HST\ and \XTE\ data, the time-resolution quoted is the nominal
precision of time-stamping; the useful time-resolution is limited by
the count rate.  No times have been given for the optical data as
these involved many short time-series, sometimes spread over several
nights.}
\label{ExpTable}
\begin{tabular}{llccccc}
\hline
\noalign{\smallskip}
Date & Instrument  & Start UT & End UT & Time       & \multicolumn{2}{c}{rms (percent)} \\ 
     &             &          &        & Resolution & Source & Noise       \\
\noalign{\smallskip}
\hline
\noalign{\smallskip}
April 6--9 
& MDM 1.3\,m, various filters & -- & -- & 6.0\,s      & --  & --            \\
\noalign{\smallskip}
April 8  
& {\it HST}/STIS E140M/1425 & 12:28:22 & 13:04:12 & 125\,$\mu$s & 3.6 & 2.1 \\
& {\it HST}/STIS E230M/1978 & 13:46:08 & 14:07:48 & 125\,$\mu$s & 4.5 & 3.3 \\
& {\it HST}/STIS E230M/2707 & 14:15:19 & 14:35:19 & 125\,$\mu$s & 4.7 & 5.3 \\
& {\it HST}/STIS E140M/1425 & 15:22:40 & 15:39:20 & 125\,$\mu$s & 3.5 & 2.2 \\
& {\it HST}/STIS E140M/1425 & 16:59:11 & 17:49:11 & 125\,$\mu$s & 3.5 & 2.1 \\
& {\it HST}/STIS E140M/1425 & 18:35:42 & 19:25:42 & 125\,$\mu$s & 3.8 & 2.1 \\
& {\it RXTE}/PCA            & 13:05:03 & 13:56:15 &  31\,$\mu$s &37.7 & 4.8 \\
\noalign{\smallskip}
April 18 
& {\it HST}/STIS E230M/1978 & 13:40:17 & 13:56:57 & 125\,$\mu$s & 4.1 & 3.3 \\
& {\it HST}/STIS E230M/2707 & 14:04:28 & 14:16:08 & 125\,$\mu$s & 4.3 & 2.4 \\
& {\it HST}/STIS E140M/1425 & 16:53:50 & 17:43:50 & 125\,$\mu$s & 3.0 & 2.1 \\
& {\it RXTE}/PCA            & 12:29:19 & 13:24:48 &  31\,$\mu$s &36.3 & 4.6 \\
& {\it RXTE}/PCA            & 14:14:23 & 18:13:19 &  31\,$\mu$s &36.5 & 4.6 \\
\noalign{\smallskip}
April 19 
& MDM 1.3\,m, various filters & -- & -- & 6.0\,s      & --  & --            \\
\noalign{\smallskip}
April 29
& {\it HST}/STIS E230M/1978 & 04:06:19 & 04:19:59 & 125\,$\mu$s & 4.3 & 3.1  \\
& {\it HST}/STIS E230M/2707 & 04:27:30 & 04:39:10 & 125\,$\mu$s & 4.1 & 2.3  \\
& {\it HST}/STIS E140M/1425 & 05:42:50 & 06:09:50 & 125\,$\mu$s & 3.6 & 2.2  \\
& {\it RXTE}/PCA            & 03:32:31 & 06:10:24 &  31\,$\mu$s &37.2 & 4.8  \\
\noalign{\smallskip}
April 30 -- May 2
& CBA-East 66\,cm, white light       & -- & -- & 3.1\,s      & --  & --           \\
\noalign{\smallskip}
May 15 
& CBA-East 66\,cm, white light       & -- & -- & 3.1\,s      & --  & --           \\
\noalign{\smallskip}
May 26 
& CBA-East 66\,cm, white light       &  -- & --       & 3.1\,s      & --  & --           \\
\noalign{\smallskip}
May 28
& {\it HST}/STIS E230M/1978 & 17:38:15 & 17:54:55 & 125\,$\mu$s & 4.2 & 3.4 \\
& {\it HST}/STIS E230M/2707 & 18:02:26 & 18:14:56 & 125\,$\mu$s & 4.6 & 2.6 \\
& {\it HST}/STIS E140M/1425 & 19:00:29 & 19:30:29 & 125\,$\mu$s & 3.3 & 2.3 \\
& {\it RXTE}/PCA            & 18:02:23 & 20:44:00 &  31\,$\mu$s &33.2 & 5.9 \\
\noalign{\smallskip}
June 24
& UKIRT/IRCAM, $K$ band     & 05:43:23 & 06:20:35 & 5--10\,s & 17.2 & 1.5\\
\noalign{\smallskip}
June 24--25
& {\it HST}/STIS E230M/1978 & 21:25:19 & 21:41:59 & 125\,$\mu$s & 3.0 & 3.2 \\
& {\it HST}/STIS E230M/2707 & 21:49:30 & 22:01:10 & 125\,$\mu$s & 3.3 & 2.5 \\
& {\it HST}/STIS E140M/1425 & 22:47:25 & 23:16:35 & 125\,$\mu$s & 2.4 & 2.4 \\
& {\it HST}/STIS E140M/1425 & 00:24:56 & 01:14:56 & 125\,$\mu$s & --  & --  \\
& {\it RXTE}/PCA            & 21:43:59 & 22:30:55 &  31\,$\mu$s &27.2 & 6.6 \\
& {\it RXTE}/PCA            & 23:19:59 & 01:32:15 &  31\,$\mu$s &27.8 & 6.2 \\
\noalign{\smallskip}
July 8
& {\it HST}/STIS E230M/1978 & 18:36:60 & 18:52:50 & 125\,$\mu$s & 3.3 & 3.7  \\
& {\it HST}/STIS E230M/2707 & 19:39:44 & 19:52:14 & 125\,$\mu$s & 3.4 & 2.9  \\
& {\it HST}/STIS E140M/1425 & 20:00:25 & 20:28:45 & 125\,$\mu$s & 2.7 & 2.5  \\
& {\it RXTE}/PCA            & 18:55:27 & 19:31:43 &  31\,$\mu$s &26.4 & 6.6  \\
& {\it RXTE}/PCA            & 20:29:03 & 21:07:21 &  31\,$\mu$s &26.3 & 6.4  \\
\noalign{\smallskip}
July 15
& UKIRT/IRCAM, $K$ band     & 05:59:22 & 06:45:28 & 5--10\,s    &48.4 & 6.7 \\
\noalign{\smallskip}
\hline
\end{tabular}
\end{table*}
\subsection{Context}
\begin{figure}
\hspace*{-5mm}\epsfig{angle=90,width=3.6in,file=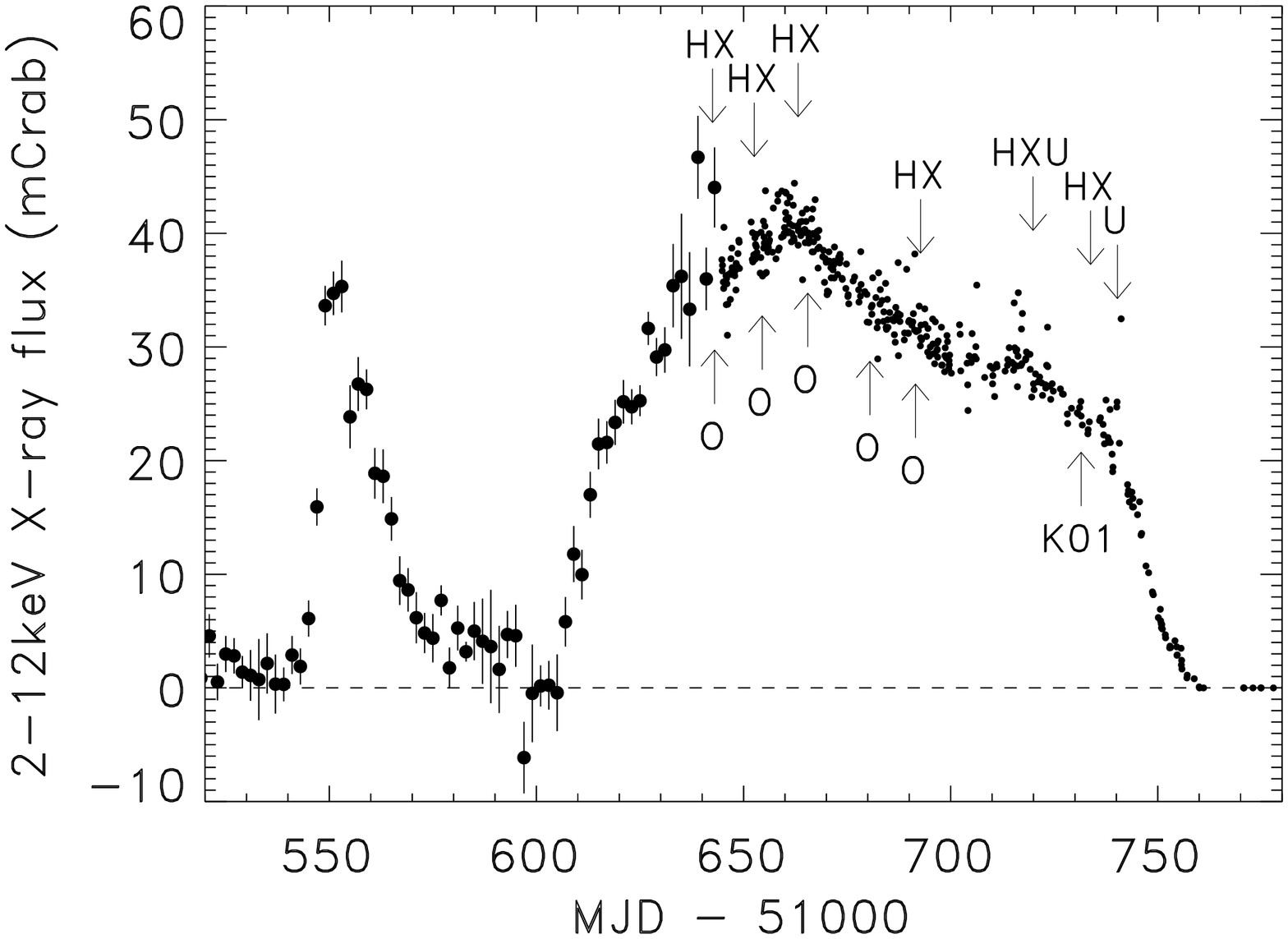}
\caption{Outburst X-ray light curve based on quick-look results
provided by the ASM/\RXTE\ team (large circles) and the {\it
ARGOS}/USA lightcurve (small circles), reproduced from Wood et al.\
(2001).  ASM data shown are two-day averages; for clarity, two-day
intervals containing less than five good dwells are omitted, and ASM
points are only shown before MJD 51644.8 when there was no USA
coverage.  Times of our observations are marked with arrows, annotated
H (\HST), X (\RXTE), U (UKIRT), or O (optical).  The times of the
optical observations of Kanbach et al.\ (2001) are also marked (K01).
These are contemporaneous with our last \HST/\XTE\ visit, but the
latter did not include any {\em simultaneous} \HST/\XTE\
observations.}
\label{LCFig}
\end{figure}

We reproduce the outburst lightcurve in Fig.\ \ref{LCFig}.  The times
of our time-resolved observations (2000 April 8 to July 15) are marked
here and summarised in Table \ref{ExpTable}.  A further observation
was carried out beyond the end of the range plotted when the source
was in or near quiescence (see Chaty et al.\ 2003) and no useful
timing information was obtained.  These observations sample the
plateau fairly well, with the first points catching the last stages of
the rise and the last points at the very end of the plateau.
\subsection{{\it HST} ultraviolet data}
\label{HSTSection}
{\it HST} observed XTE\,J1118+480 using the Space Telescope Imaging
Spectrograph (STIS) on seven visits spanning 2000 April--September.
All UV observations were obtained in TIMETAG mode, giving a record of
arrival time of individual photons with a relative precision of
125\,$\mu$s.  A log of the useful exposures is given in Table
\ref{ExpTable}; the count rate for the September visit was too low to
be useful for a timing analysis.

Three observing modes were used: E140M/1425, a medium
resolution echelle grating with central wavelength 1425\,\AA,
E230M/1978, and E230M/2707, which are defined similarly.  The E140M
mode used the far-UV MAMA while the E230M modes used the near-UV MAMA.
For convenience we will sometimes refer to these three modes as
far-UV, mid-UV, and near-UV respectively, although these are not
standard definitions.  We find typical global count rates of
1600--2200\cntpersec, 2700--3300\cntpersec\ and 3300--4300\cntpersec\
for the E140M/1425, E230M/1978 and E230M/2707 modes respectively.  The
corresponding background rates (dominated by the instrumental dark
current) are 7\cntpersec\ for E140M and 1100--1500 for E230M modes
(Leitherer et al.\ 2001, Proffitt, C., 2001, priv.\ comm.).
Geocoronal emission lines are detected by the E140M, but their
contribution is negligible compared to source counts and dark current.
Clearly the E140M background, at a level of $<0.5$\,percent is
negligible, so we simply subtract an average value of 7\cntpersec\
from these data.  For the E230M modes, however, the large dark current
of the STIS NUV MAMA (due to phosphorescence in the detector faceplate
window; Ferguson \& Baum 1999) does need to be accounted for more
carefully, especially as it varies with detector temperature.  The
echelle gratings leave no good unexposed region to measure it because
there is a significant amount of scattered light between the orders,
but an approximate model is available (Leitherer et al.\ 2001,
Proffitt, C., 2001, priv.\ comm.), and we used this model to dark
subtract the E230M lightcurves.  Fortunately, for this study we are
most interested in high frequencies which should not be contaminated
by residual variations in the dark current.

The first observation on June 24--25 shows a significant, but
declining, excess in the count rate for the first 200\,s.  The
distribution of the excess events across the detector is consistent
with it being due to a high dark current; it is likely that this is
due to the temperature changing rapidly just after the MAMA was turned
on for the day (Proffitt, C., 2001, priv.\ comm.).  We discarded the
affected data as no reliable correction could be made and there were
no simultaneous \XTE\ data.  The last \HST\ observation on June 24--25
shows about 80\,percent of the count rate of the previous one,
significant low-frequency variations, and a large step towards the
end.  We therefore treated these data with considerable caution, but
as there were simultaneous \XTE\ data we do not completely reject
them.  We discarded the low-frequency information from this exposure,
but high-frequencies, and correlations with X-ray variations, are
likely to still be useful.

All lightcurves were barycenter corrected using the {\sc stsdas} task
{\sc odelaytime}; the corrections are consistent with those generated
for the \XTE\ lightcurves, with only small (real) differences due to
the different positions of the spacecraft.  Finally the barycenter
corrected TIMETAG events were binned in time using 1/64\,s, 1/16\,s or
1\,s bins depending on the analysis to be done.  Care was taken to
reject partially filled bins at the edge of a good observing period.

The final absolute timing accuracy is limited by the accuracy of the
\HST\ and STIS clocks as the photon time stamping (with 125\,$\mu$s
precision) and barycenter corrections are much more precise than this.
The accuracy of the STIS clock itself is poorly known (Sahu K., 2001,
priv.\ comm; Gull T., 2002, priv.\ comm.).  It is reset from the
spacecraft clock every time the satellite recovers from safe mode, but
after that it can drift.  The drift is likely to be slow but is not
known.  If the drift is linear, an upper limit can be placed from the
derived UV period of the Crab pulsar (Gull et al.\ 1998).  The UV
period agrees with the radio one to better than 1\,ns, implying a
clock drift of less than 1 part in $3.3\times10^7$, or
2.6\,ms\,day$^{-1}$.  Fortunately because of the gyro failure in 1999
November, there was a safe mode recovery on 1999 December 28, less
than four months before our observations begun, so the accumulated
drift should not amount to very much.  The maximum possible drift
relative to the spacecraft clock is 0.26\,s for the first visit and
0.47\,s for the last with simultaneous coverage.  This depends upon
the drift being linear, however, which is not certain.  There may also
be some drift of the spacecraft clock with respect to UTC.
Consequently, the error in the STIS absolute timing could be larger
than these estimates, possibly a number of seconds, but probably not
minutes (Long 2000).  There {\em could} then be a significant overall
offset affecting all of the STIS lightcurves.  The constraint on the
STIS drift rate from the measured Crab period, and the consistency of
the lags we subsequently measure with respect to X-rays
(Section~\ref{LagSection}) does indicate that this offset is
approximately consistent within the dataset, so that all observations
are offset by approximately the same amount.

\subsection{Optical data}

High-speed optical photometry was obtained at the 66\,cm telescope of
the Maryland observing station of the Center for Backyard Astrophysics
(``CBA-East'', described by Skillman \& Patterson 1993).  The data
consisted of several 52\,min time series in white light
(4000--8000\,\AA), with a time resolution (integration plus
readout time) of 3.1\,s.  Typically four 52\,min datasets were
obtained during a night.  Timing accuracy was maintained to $<1$\,s
during each run, but the absolute timing was known only to
$\sim2$\,s.  On one night, April 19, we also obtained photometry from
the MDM 1.3\,m using various filters at a time resolution about 6\,s.

\subsection{IR data}
Near-infrared observations of \target\ were carried out at the United
Kingdom Infrared Telescope (UKIRT) 3.8\,m telescope using the
$1024\times1024$ pixel UFTI (1--2.5\,$\umu$m) camera (pixel scale
0.09\,arcsec) and a broad band $K$ filter (2.03--2.37\,$\umu$m) on
2000 June 24 and July 15.

On June 24, we obtained 40\,min of data with between 2\,s and 4\,s
integration time, and a total of 240 good frames. The airmass was
$\sim1.35$.  On July 15, we took 1\,hr of data, a total of 299 good
frames, with an integration time of 2\,s for all the frames. The
airmass was $\sim1.7$.  The sampling time was nearly 7\,s for both
observations, which was the quickest we could achieve with UKIRT/UFTI.

To ensure accurate correction of the bright infrared sky we performed
30\,arcsec offsets to the North-West, North-East, South-East and
South-West from the central position.  In order to reduce the overhead
time, we took many exposures of the object between offsets, the number
of exposures depending on the quality of the sky.

The images were processed using {\sc iraf}\footnote{IRAF is
distributed by the National Optical Astronomy Observatories, which are
operated by the Association of Universities for Research in Astronomy,
Inc., under cooperative agreement with the National Science
Foundation.} reduction software. Each of the images were corrected for
the dark current, normalised with a flat field, and sky-subtracted by
a sky image created from median-filter combining a total of 9 (or
more) consecutive images.  The data were then analysed using the {\sc
apphot} task within {\sc iraf}, taking different apertures depending
on the seeing. Formal errors were estimated from the source brightness
and the standard deviation of the sky in the usual way.  The
conditions were photometric for most of the observations, the seeing
being typically 0.8\,arcsec.

Absolute photometric calibration was performed for both observations
using a nearby photometric standard star from the UKIRT Extended list
(Hawarden et al.\ 2001): FS 130 (P264-F), with $K$ band exposures (5
images each of 8\,s).  We derived the apparent magnitude of \target\
by co-adding and median filtering all the individual frames.  The
average magnitude was $K = 11.512 \pm 0.004$ and $K = 11.948 \pm
0.006$ on June 24 and July 15 respectively.
\subsection{X-ray data}
\label{XTESection}
\target\ was observed with the instruments aboard \RXTE\ at several
epochs simultaneous or contemporaneous with our \HST\ observations.
For this study, we only used data from the PCA, which covers a nominal
energy range of 2--60\,keV.  The number of PCUs used varied; the
lightcurves used here were constructed from intervals when 3--5\,PCUs
were in use.  The data were taken in various data modes running in
parallel.  We used standard \RXTE\ data screening and reduction
techniques to construct light curves, integrated over the 2--60\,keV
energy range, by rebinning data from several high-timing-resolution
data modes.  Barycenter corrections used the JPL DE200 ephemeris.  The
relative timing accuracy of the \RXTE\ data is limited only by the
stability of the spacecraft clock which is good to about 1\,$\mu$s or
less.  The absolute timing accuracy, however, is also limited by
uncertainties in the ground clock at the White Sands station and other
complications and is estimated to be about 5\,$\mu$s which is
substantially better than that of \HST/STIS and certainly sufficient
for our study.  As \target\ is a relatively faint source, we also
estimated background lightcurves using {\sc pcabackest} from {\sc
heasoft} 5.0 and background models dated 2000 January 31.  The
background values are only required so that the source fractional
variability can be correctly estimated, so the accurate values are not
critical.  This background $\sim30$\,cnt\,s$^{-1}$\,PCU$^{-1}$
($\sim25$\,percent of the total count rate) was subtracted from the
lightcurves prior to further analysis.  Binned lightcurves were
constructed in a similar way to the \HST\ ones, with three time
resolutions, $1/64$\,s, $1/16$\,s and 1\,s.  Source count rates were
80--120\,cnt\,s$^{-1}$\,PCU$^{-1}$.
%
%%%%%%%%%%%%%%%%%%%%%%%%%%%%%%%%%%%%%%%%%%%%%%%%%%%%%%%%%%%%%%%%%%%%%%%%%%%%%%%
%
\section{Lightcurves}
\begin{figure}
\hspace*{-5mm}\epsfig{angle=90,width=3.6in,file=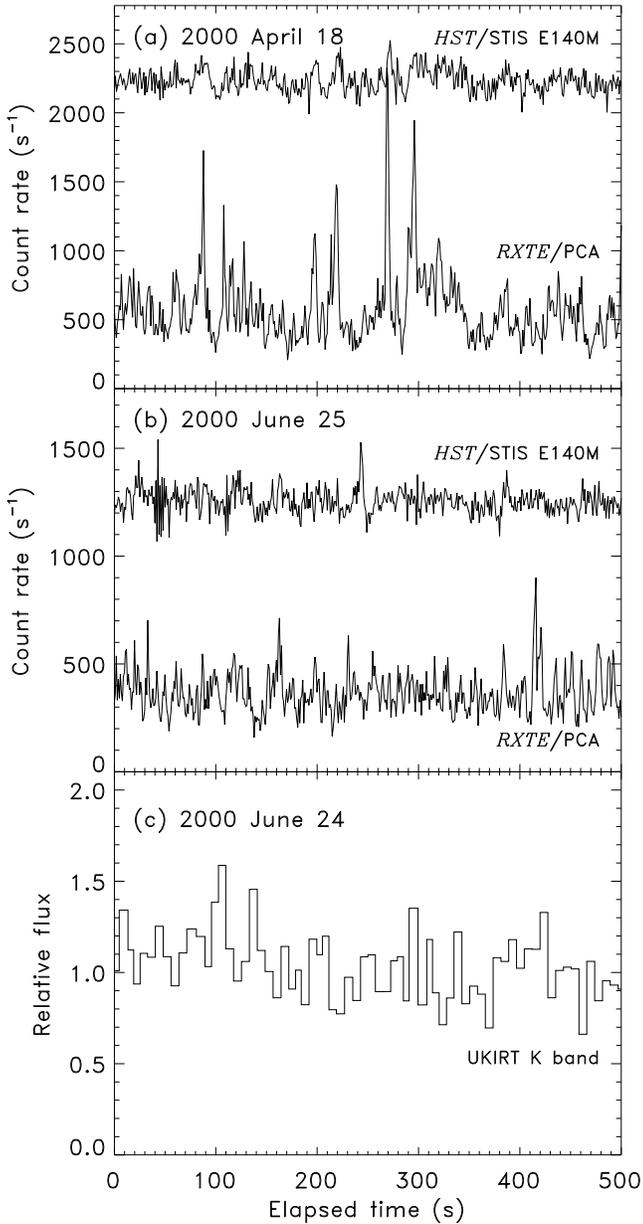}
\caption{a,b) Examples of simultaneous \RXTE\ and \HST\ lightcurves.
Note that the strongest features in the X-ray lightcurve sometimes
show clearly correlated features in the \HST\ UV lightcurve.  This is
especially clear in (a).  At other times, no obvious UV counterpart is
present.  The lower amplitude variability at the later epoch reflects
the changing PSD.  c) A segment of an IR $K$ band lightcurve obtained
from UKIRT on 2000 June 24, contemporaneous with panel (b).  The
timespan covered is the same as that in Fig.\ \ref{RapidLCFig} for
comparison.  The error bars are only 1.5\,percent so are negligible
compared to the observed variations.  The amplitude of the variability
is clearly closer to that in the X-rays than in the UV.}
\label{RapidLCFig}
\end{figure}

\target\ exhibited dramatic short term variability at all wavelengths
with sufficient signal-to-noise to detect it.  The variability is
particularly dramatic at X-ray energies where flares can rise to
$5\times$ the baseline flux in a few seconds (Fig.\ \ref{RapidLCFig}a)
and in the infrared (Fig.\ \ref{RapidLCFig}c).  Ultraviolet
lightcurves exhibit similar, although more subdued variability which
is clearly correlated with the X-ray flares (e.g.\ Fig.\
\ref{RapidLCFig}a).  Optical variability is also present with
amplitude intermediate between the UV and IR, and is correlated with
X-ray variations (Kanbach et al.\ 2001).

In Table~\ref{ExpTable} we have summarised the intrinsic and noise rms
of the lightcurves.  For the \HST\ and \XTE\ lightcurves these are
based on 1\,s binned lightcurves, so indicate the variability on
timescales longer than this.  For the UKIRT data they are based on the
raw time resolution of the data, 5--10\,s, so exclude power above
$\sim0.1-0.2$\,Hz.  Consequently the UKIRT source rms values quoted
are underestimates compared with the others.  The strength of the
X-ray variability (25--40\,percent) appears to decrease with time and
is most dramatic in the 2000 April observations (near the peak of the
\RXTE/ASM lightcurve in Fig.\ \ref{LCFig}).  The UV variability at a
2--5\,percent level also decreases with time, but increases with
wavelength.  The same trend appears to extrapolate into the optical
and IR: the typical scatter in our optical data is about 20\,percent;
Kanbach et al.\ (2001) measure a broad band optical rms
$\sim10$\,percent later in the outburst; and the UKIRT data indicate
an rms on comparable timescales of $\ga17$\,percent for the June
observation and $\ga50$\,percent for July.  This trend already
suggests that the variability is not due to disc reprocessing (which
should be blue) but rather appears to be veiled by a less variable
blue disc component.  It is probably due to a non-thermal mechanism,
likely synchrotron as previously suggested by Merloni et al.\ (2000)
and Kanbach et al.\ (2001).  This interpretation would be consistent
with the continuity of the spectral energy distribution from the
optical to the radio (Hynes et al.\ 2000; Chaty et al.\ 2003).

\begin{figure}
\hspace*{-5mm}\epsfig{angle=90,width=3.6in,file=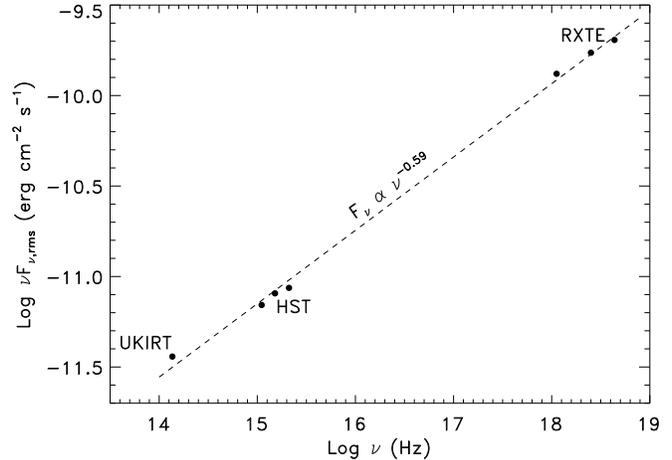}
\caption{Broad-band spectral energy distribution of the rms
  variability on 2000 June 24--25.  Note that the IR data not obtained
  simultaneously with the other bands, and that the plotted value
  actually represents a lower-limit, since these data were obtained at
  lower time-resolution.}
\label{RMSSEDFig}
\end{figure}

It is useful to use these rms values to to examine the broad-band
energy distribution of the variability more quantitatively; this is
effectively an rms spectral energy distribution.  To construct this we
use the rms values from Table~\ref{ExpTable} together with the mean
flux in each band to absolutely calibrate the rms values.  We use the
data from 2000 June 24--25 when we have both IR and UV coverage.  This
gives us estimates at $K$, 2707, 1978, and 1425\,\AA\, and in the
X-rays.  To examine the X-ray data in more detail we extracted
lightcurves for 2--7.5, 7.5--15.2, and 15.2--60\,keV and measured the
noise-subtracted rms, after subtracting the interpolated background
level.  The resulting energy distribution is shown in
Fig.~\ref{RMSSEDFig}.

It has a surprisingly simple form, and can be well represented
by a single power-law, $f_{\nu} \propto \nu^{-0.59}$.  This is
different to the mean spectrum (Hynes et al.\ 2000; Chaty et al.\
2003), for which the IR--UV is essentially flat ($f_{\nu} =
$\,constant).  The slope would be consistent with optically thin
synchrotron emission; we will return to discussion of this possibility
in Section~\ref{DiscussionSection}.
%
%%%%%%%%%%%%%%%%%%%%%%%%%%%%%%%%%%%%%%%%%%%%%%%%%%%%%%%%%%%%%%%%%%%%%%%%%%%%%%%
%
\section{Variability Properties}
\subsection{Autocorrelation functions}
\label{ACFSection}
We begin an analysis of the variability properties of the lightcurves
by constructing autocorrelation functions (ACFs).  This is useful both
for a direct comparison with the optical results of Kanbach et al.\
(2001), and also because the white-noise contribution is better
isolated in an ACF than in the PSD.

\begin{figure*}
\epsfig{angle=90,width=7in,file=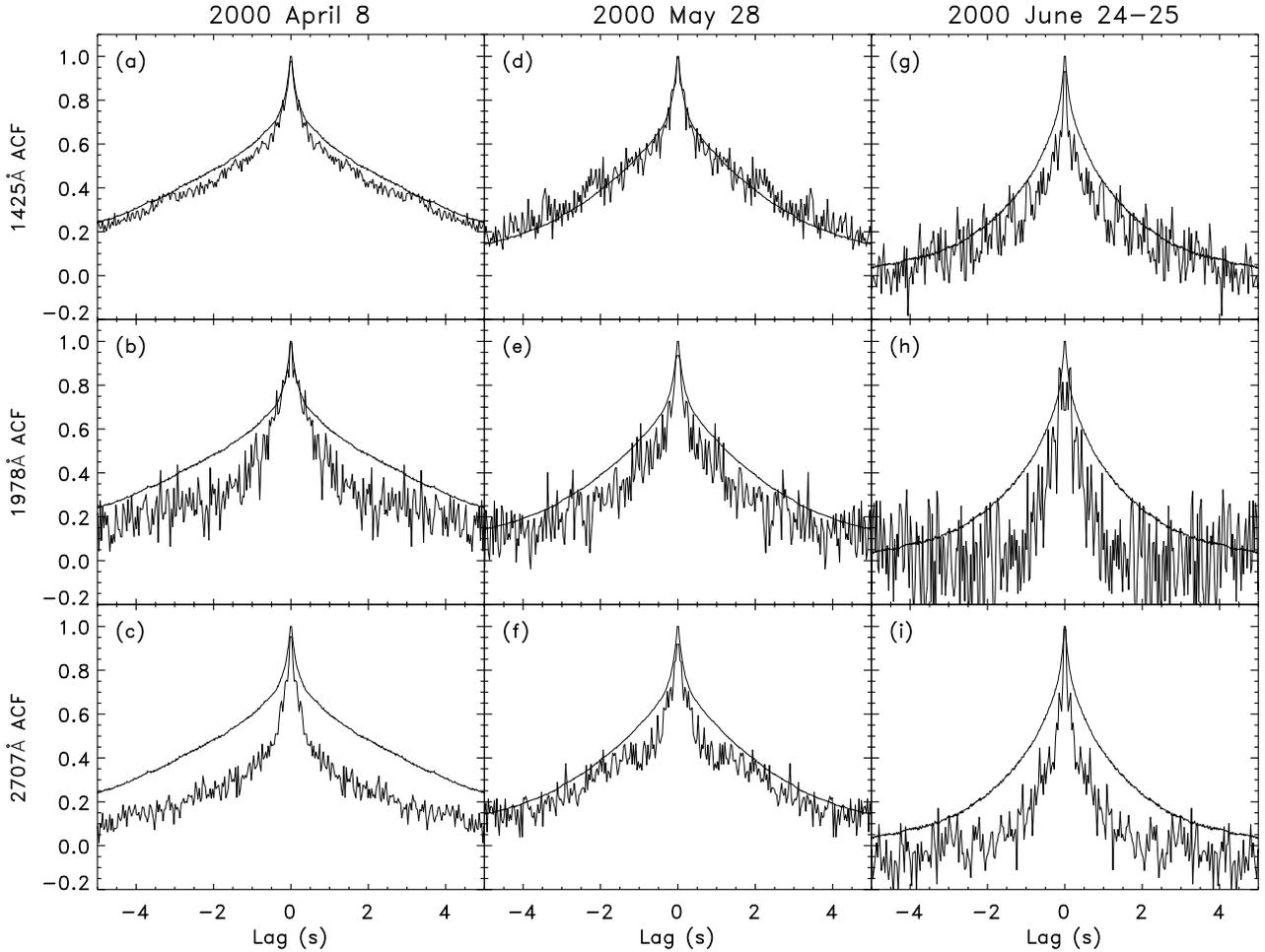}
\caption{A selection of X-ray and ultraviolet autocorrelation
functions.  In each panel, the smoother ACF is the X-ray one and the
noisy one is UV.  The zero lag bin (containing noise) was clipped out.
Each ACF then had the mean response in the 30--50\,s range subtracted
to remove the contribution from very low frequencies, then was
normalised to a peak {\em unbinned} value of unity.  Each ACF is the
average of all the data in that band for each epoch, i.e.\ no attempt
has been made to show ACFs from simultaneous data.  Note the tendency
for the short-wavelength ACF to look more like the X-ray one, with a
sharper component becoming stronger at long wavelengths.  The X-ray
ACF itself becomes sharper with time as the PSD evolves.}
\label{ACFFig}
\end{figure*}

Since our \HST\ and \XTE\ lightcurves are uniformly binned with no
gaps this is straightforward and no interpolation is necessary.  An
ACF is especially appealing as Poisson noise will have no
autocorrelation and so its ACF is a delta function at zero lag.  We
therefore use the high-resolution lightcurves ($1/64$\,s) and exclude
the zero lag point to remove the contribution of Poisson noise to the
ACF.  For each visit we construct average ACFs for X-rays and each UV
band (central wavelengths 1425, 1978, and 2707\,\AA).  A selection are
plotted in Fig.~\ref{ACFFig}, with each panel comparing one UV ACF
with the contemporaneous X-ray ACF.  Fig.~\ref{ACFFig}a--c shows data
from 2000 April 8; this has the best UV coverage so best illustrates
the wavelength dependence of the ACFs.  It is clear that the far-UV
ACF is almost identical to the X-ray one, but has a slightly sharper
peak.  At longer wavelengths the sharpness of the peak increases and
the near-UV ACF is dramatically different to the X-ray one.  The sharp
central peak is strikingly similar to the optical ACF shown by Kanbach
et al.\ (2001), with a width of $<1$\,s.  The same trend is seen in
all of the data with narrower ACFs at longer wavelengths.
Fig.~\ref{ACFFig}d--f shows data from 2000 May 28, the visit showing
the most similarities between X-ray and UV ACFs, and
Fig.~\ref{ACFFig}g--i shows data from 2000 June 24--25 which
illustrates the sharply peaked ACFs most dramatically, and shows that
at times even the far-UV can be sharply peaked.

As Kanbach et al.\ (2001) pointed out, an ACF narrower than the X-ray
one is hard to reconcile with reprocessing; the smearing due to
varying light travel times should always broaden the ACF.
Furthermore, one would expect the longer wavelengths to originate
preferentially from cooler regions further from the X-ray source;
hence they should suffer more smearing and have broader ACFs, not
narrower ones.  The problem can be overcome if the UV response does
not vary linearly with the X-ray irradiation.  While a non-linear
response might be expected in lines, it is harder to acheive in the
continuum.  In a simple irradiated black body model the response can
be non-linear (on the Wien part of the black body spectrum), but the
effect is most pronounced at shorter wavelengths, which is also the
opposite of what is observed.  Another way to distort the ACFs is if
the response is sometimes negative.  We will discuss non-linear and
negative responses further in Sections~\ref{NonLinearSection} and
\ref{NegativeSection} respectively; neither of these effects can
explain the observations in the context of a reprocessing model.

There are two ways to interpret the wavelength dependence of the ACF.
Either there is a single variable component with wavelength dependent
properties (presumably synchrotron as the variability extends to the
IR), or there are two components, a weak broad response seen at short
wavelengths and a sharper, higher amplitude component dominant at
longer wavelengths.  A natural interpretation for a two component
model is provided by our decomposition of the SED (Hynes et al.\ 2000;
Chaty et al.\ 2003).  It is likely that the UV is dominated by the
accretion disc (though with some synchrotron contribution) and that
the IR is almost pure synchrotron.  It is then natural to suggest that
the broad component of variability dominating in the far-UV is disc
reprocessing but that the IR, optical and near-UV variability is due
to synchrotron emission.  We will address this question further in the
following sections and attempt to discriminate between these
possibilities.

It would be useful to have a comparable IR ACF.  Unfortunately the
UKIRT data have insufficient time-resolution for this.  We have
constructed ACFs using the discrete correlation function method of
Edelson \& Krolik (1988), but find essentially no autocorrelation
outside the zero-lag bin in either UKIRT observation.  This indicates
that no significant autocorrelation is present at lags $\ga5$\,s.
Since X-ray, UV and optical ACFs also show little autocorrelation at
$\ga5$\,s, however, this is not a strong constraint.

\subsection{Broad band power spectra}
\label{PowerSpecSection}
To further characterise the variability we have constructed power
spectral densities (PSDs).  We perform a Fourier transform of each
uninterrupted lightcurve segment (in many cases a whole lightcurve of
several thousand seconds to preserve low frequency information).  We
use a variable sized binning in frequency to produce the PSDs shown in
the figures.  The constraints imposed on the binning are: i) a minimum
and maximum logarithmic bin width, ii) a minimum signal-to-noise ratio
and iii) a minimum of three points per bin.  Within each bin, we use
the standard deviation of the points to estimate the error on the
binned power.  This is obviously unreliable for the lowest frequencies
($\la0.003$\,Hz), so we exclude these from the fits.  Where
appropriate, PSDs from several lightcurve segments have been combined
before this binning, particularly for the X-ray data.  PSDs have been
plotted with the white noise contribution subtracted and normalised to
fractional rms amplitude squared per hertz (c.f.\ van der Klis 1995).
The white noise level was determined empirically by fitting a red
noise plus white noise model to the high frequency data.  The optical
time-resolution was not sufficient for this approach, so no white
noise subtraction has been done for that.

\begin{figure}
\hspace*{-5mm}\epsfig{angle=90,width=3.6in,file=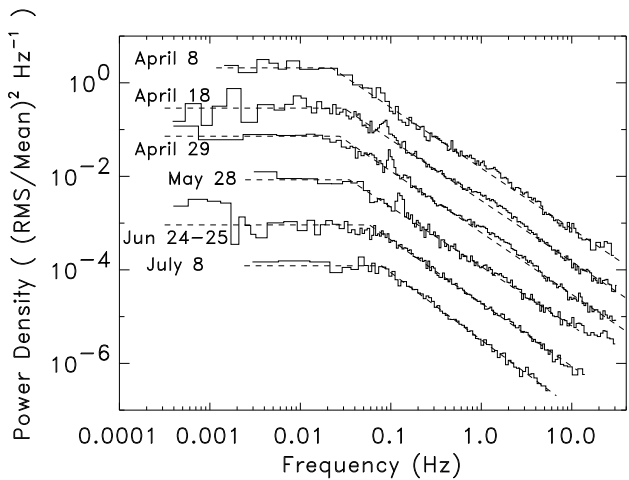}
\caption{Average \RXTE\ power spectrum from each visit.  The
statistical white noise contribution has been subtracted from each PSD
and there is a downward offset of a factor of 5 between each
successive PSD.  The fit is described in the text.  Note the QPO
evolution from April 18 to May 28 and the disappearance of the high
frequency bump at later epochs.}
\label{PSDFig}
\end{figure}

\begin{figure}
\hspace*{-5mm}\epsfig{angle=90,width=3.6in,file=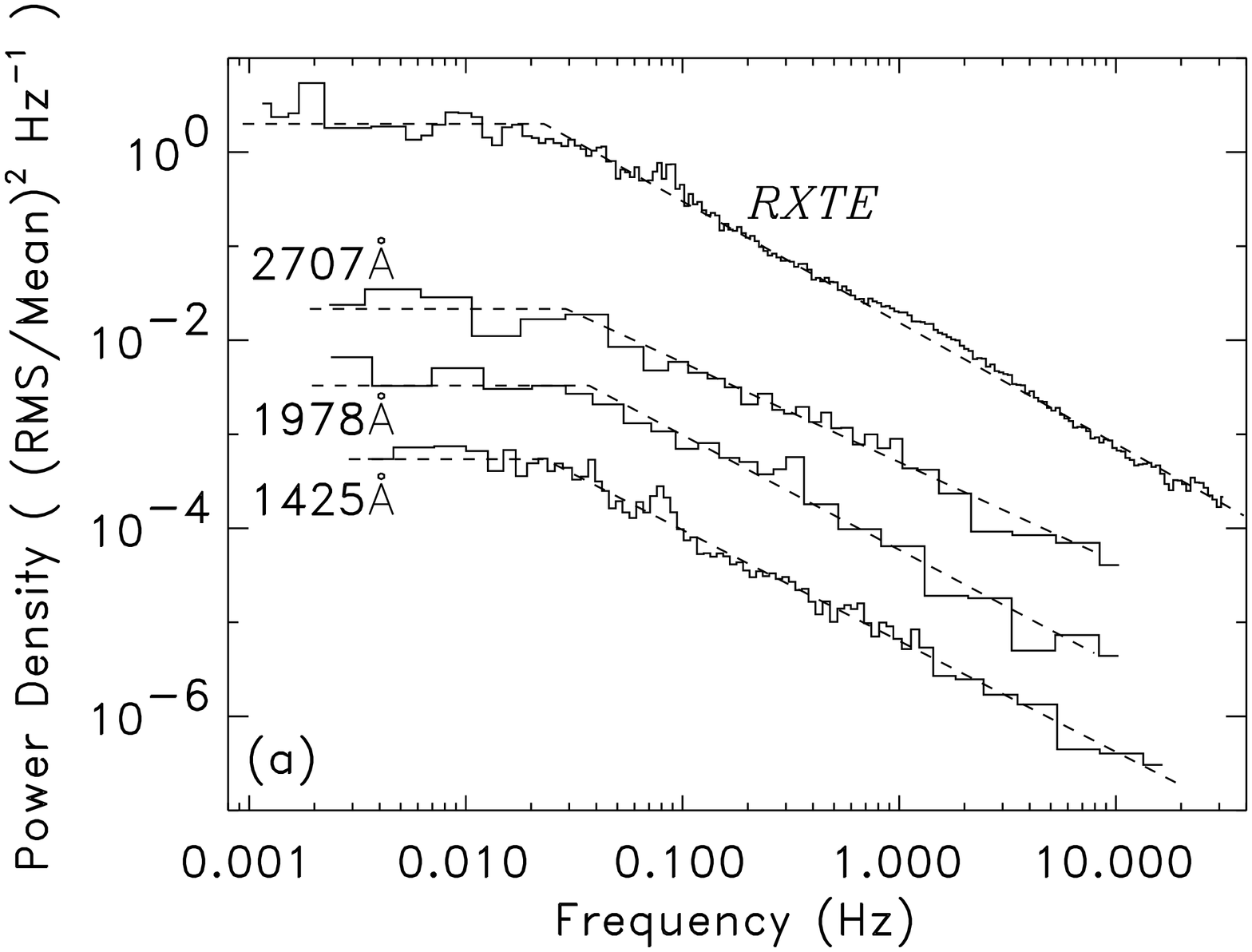}
\hspace*{-5mm}\epsfig{angle=90,width=3.6in,file=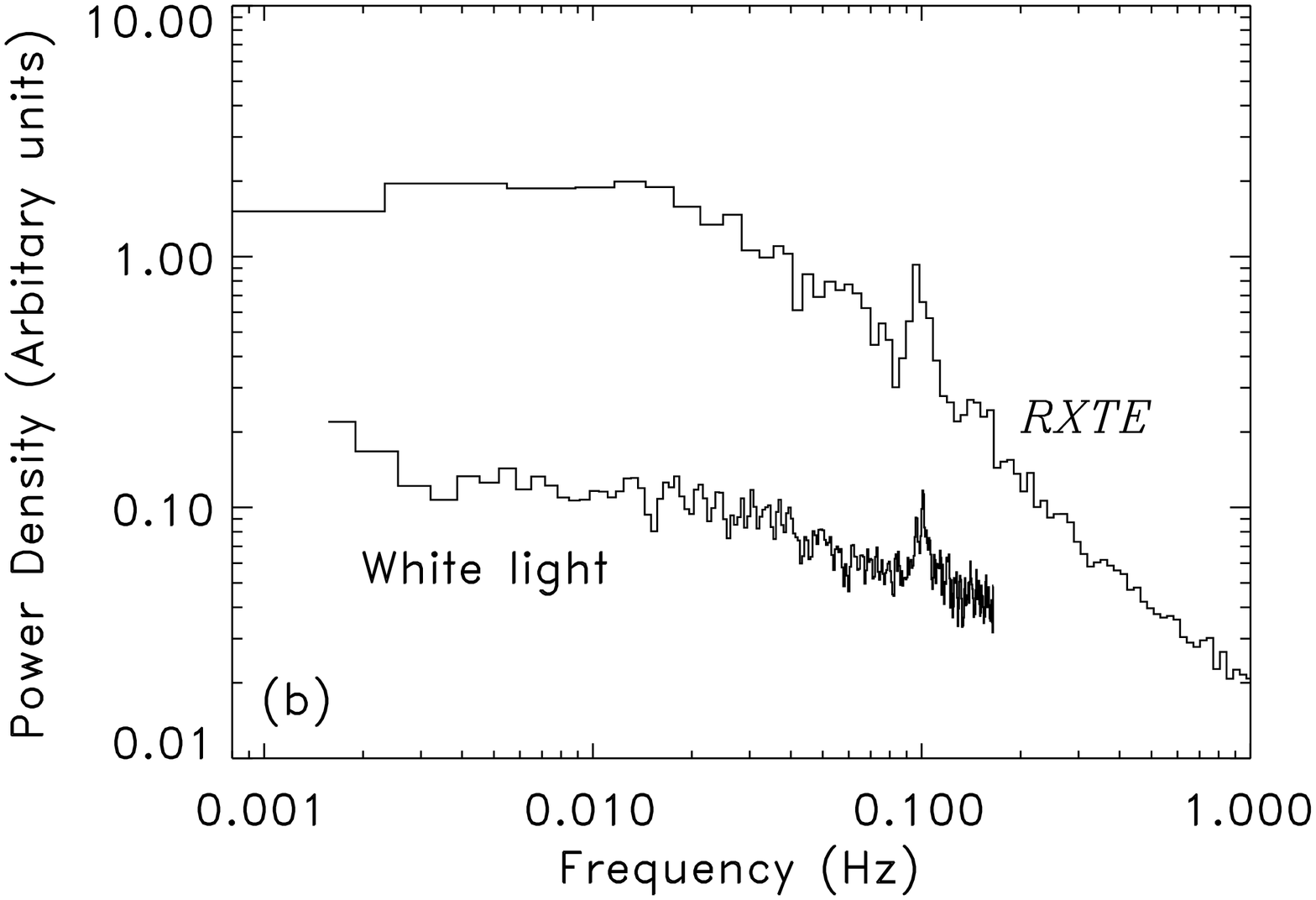}
\caption{a) X-ray and UV PSDs from the first and second visits.  The
\XTE\ and 2707\,\AA\ PSD is normalised correctly; the 1978\,\AA\ and
1425\,\AA\ ones have been offset downward by factors of 5 and 25
respectively.  Note that the data are not all strictly simultaneous.
b) X-ray and optical PSDs from around the 3rd visit.  Again these are
not strictly simultaneous; the optical PSD is the average of 40
12.5\,min runs during April 30--May 2. The \XTE\ data are normalised
as in a), but the optical normalisation is arbitary.  The
high-frequency white noise contribution has not been removed from the
optical data.  Note the similarity of the QPOs.}
\label{MWPSDFig}
\end{figure}

As noted by Revnivtsev, Sunyaev \& Borozdin (2000), \RXTE\ PSDs of
\target\ closely resemble power-spectra of other black hole candidates
in the low-state (Wijnands \& van der Klis 1999), exhibiting a flat
spectrum at low frequencies and a declining power-law at higher
frequencies (Fig.\ \ref{PSDFig}).  All of the \HST\ UV PSDs are
similar (Fig.~\ref{MWPSDFig}), but with a lower amplitude, as are the
optical ones.  IR PSDs exhibit a white noise structure with an
amplitude similar to, but slightly higher than, the X-ray data.  The
IR PSD does not extend to high frequencies, and unfortunately coverage
stops around the break frequency so we cannot tell if the break is
present in the IR.  QPOs also seem to be present in X-ray, UV, and
optical data.  These are discussed in Section~\ref{QPOSection}.

To better characterise the temporal evolution and wavelength
dependence of the PSD, we fit them (after subtracting the white noise
level which dominates at high frequencies) with a simple model
comprising a red-noise power-law above a variable break frequency and
white noise below this.  Parameters of interest to be fit are the
break frequency and the red noise amplitude and slope.  Regions
containing the QPO or a pronounced high frequency bump are masked out
from the fits.  As the outburst proceeds we find the break frequency
moving to higher frequencies, in a similar way to the QPO, and the
high-frequency bump (visible at 1--5\,Hz in the first three visits;
Fig.~\ref{PSDFig}) disappears.  The red noise power and slope,
however, show little variation.  As a function of wavelength, UV PSDs
have a very similar shape to the X-ray ones, but appear to break at a
slightly higher frequency and have a flatter slope.  The UV amplitude
is higher at longer wavelengths, as already deduced from the overall
rms.

\begin{table}
\caption{Fits to \RXTE\ PSDs as a function of epoch.  The model is a
broken power law.  The quoted power and slope define the power law.
Below the break frequency the model is flat.  The QPO and broad high
frequency bump are masked out of the fits.}
\label{PSDTable}
\begin{tabular}{lccc}
\hline
\noalign{\smallskip}
Date             & Break     & Power & Slope \\
                 & frequency & at 0.1\,Hz  &       \\ 
\noalign{\smallskip}
2000 April 8     & 0.022     & 0.30        & -1.30 \\
2000 April 18    & 0.030     & 0.30        & -1.30 \\
2000 April 29    & 0.027     & 0.32        & -1.32 \\
2000 May 28      & 0.035     & 0.28        & -1.27 \\
2000 June 24--25 & 0.058     & 0.27        & -1.35 \\
2000 July 8      & 0.079     & 0.27        & -1.44 \\
\noalign{\smallskip}
\hline
\end{tabular}
\end{table}

\begin{table}
\caption{Fits to \RXTE\ and \HST\ PSDs as a function of wavelength for
the first two epochs.  Details as for Table~\ref{PSDTable}.}
\label{HSTPSDTable}
\begin{tabular}{lccc}
\hline
\noalign{\smallskip}
Date             & Break     & Power & Slope \\
                 & frequency & at 0.1\,Hz  &       \\ 
\noalign{\smallskip}
\RXTE/PCA             & 0.023     & 0.30          & -1.29 \\
\HST/STIS, 1425\,\AA\ & 0.023     & 0.0024        & -1.18 \\
\HST/STIS, 1978\,\AA\ & 0.037     & 0.0049        & -1.22 \\
\HST/STIS, 2707\,\AA\ & 0.029     & 0.0058        & -1.06 \\
\noalign{\smallskip}
\hline
\end{tabular}
\end{table}

\subsection{The quasi-periodic oscillation}
\label{QPOSection}
\target\ exhibits a prominent low-frequency QPO present in X-rays
(Revnivtsev et al.\ 2000) and also in the UV and optical (Haswell et
al.\ 2000).  Wood et al.\ (2000) have performed a thorough analysis of
\RXTE/PCA and {\it ARGOS}/USA X-ray data.  They find that the X-ray
QPO is detectable for a period of about two months during the
outburst, with an amplitude of 5--10\,percent rms and a frequency
increasing monotonically from 0.07\,Hz to 0.16\,Hz.  We have examined
all of our data to compare the multi-wavelength properties of the QPO.
Wood et al.\ (2000) only presented data extending up to 2000 June 11.
We have two epochs of observations after this (June 24--25, July 8),
but find no detectable QPO in these visits.

\begin{figure}
\hspace*{-5mm}\epsfig{angle=90,width=3.6in,file=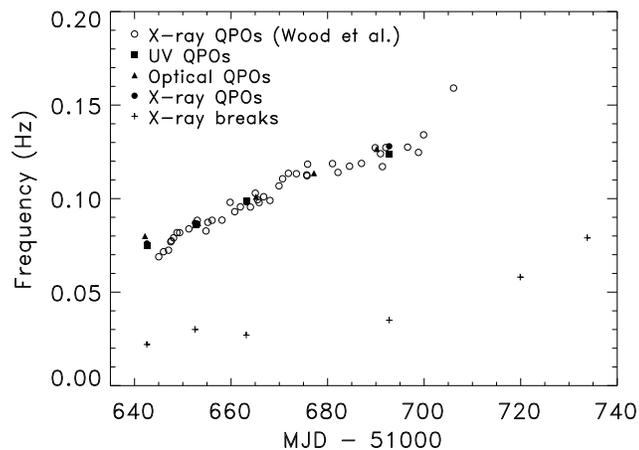}
\caption{Evolution of QPO and break frequencies with time.  Open
circles points are QPO measurements derived from Wood et al.\ (2000).
Other points are based on measurements described herein.}
\label{QPOFig}
\end{figure}

We performed the QPO search rather differently to the analysis of the
broad band power.  We divided each \RXTE\ and \HST\ lightcurve up into
256\,s segments and calculated an FFT of each segment.  We then
averaged all the \RXTE\ and HST FFTs from each visit to construct
X-ray and UV power-spectra.  This approach should improve the
sensitivity to low-coherence features.  We fitted the region around
the QPO ($\sim \pm 0.05$\,Hz) with a model comprising a red noise
power-law and a Lorentzian QPO.  We summarise the fitted QPO
frequencies, FWHM and integrated amplitudes of the model fit in
Table~\ref{QPOTable}.  Similarly, we divided each optical run into
10\,minute segments, and averaged the 20--40 individual power spectra.
We also list the frequencies of QPOs detected in the optical data in
Table~\ref{QPOTable}.  The UV and optical QPO frequencies are the same
as the X-ray ones, with some frequency evolution as shown by Wood et
al.\ (2000).  We note that X-ray, UV, and optical frequencies from
April 8 are all consistent and significantly different to the
frequencies measured a few days later.  This short-term variation in
the QPO frequency therefore appears be a real effect.  In
Fig.~\ref{QPOFig} we show both the frequencies reported by Wood et
al.\ (2001) and our own measurements.  The fractional rms of the X-ray
QPO is 8--10\,percent, consistent with earlier reports.  The UV QPOs
have an amplitude of only $\sim1$\,percent.  Both the X-ray and UV
QPOs account for $\la10$\,percent of the total variance below 1\,Hz,
and so should not dominate the lightcurves or correlation studies.
There is no evidence that the UV QPO is stronger or weaker relative to
the aperiodic variability than is the X-ray QPO; neither does it
appear to be systematically narrower or broader.  Consequently it is
likely that the same mechanism as produces correlated multi-wavelength
aperiodic variability are also responsible for the correlated QPO.
With a QPO frequency 0.08--0.13\,Hz and a break frequency
0.02--0.04\,Hz (for the first four visits), the PSDs are approximately
consistent with with the relation between the frequencies shown by
Wijnands \& van der Klis (1999) for other sources, although both
frequencies are somewhat below the lowest values collated by those
authors.

\begin{table}
\caption{Fitted properties of the low-frequency QPO.  Each measurement
was obtained from the average of all the PSDs from the same visit and
telescope.  For \HST\ data, we have averaged the near- and far-UV data
together to improve the signal-to-noise; the QPO is clearly present in
both.  QPOs were not detected in the June or July visits.  Only the
frequency was determined for the optical QPOs.  See
Table~\ref{ExpTable} for further details of the data.}
\label{QPOTable}
\begin{tabular}{llccc}
\hline
\noalign{\smallskip}
Date & Bandpass & Freq.\ & FWHM & rms       \\
     &           & (Hz)      & (Hz) & (percent) \\ 
\noalign{\smallskip}
2000 April 8  & UV   & 0.075 & 0.018 & 0.9 \\
              & X-ray   & 0.076 & 0.004 & 8.0\\
              & Optical & 0.080 & --    & --  \\
\noalign{\smallskip}
2000 April 18 & UV   & 0.086  & 0.016 & 1.0 \\
              & X-ray   & 0.087  & 0.016 & 9.5 \\
\noalign{\smallskip}
2000 April 19 & Optical & 0.087  & --    & --  \\
\noalign{\smallskip}
2000 April 29 & UV & 0.099  & 0.013 & 1.1 \\
              & X-ray & 0.098  & 0.007 & 8.5 \\
\noalign{\smallskip}
2000 May 1 & Optical & 0.101   & --    & --  \\
\noalign{\smallskip}
2000 May 15 & Optical & 0.114  & --    & --  \\
\noalign{\smallskip}
2000 May 26 & Optical & 0.127  & --    & --  \\
\noalign{\smallskip}
2000 May 28   & UV & 0.124  & 0.015 & 1.1\\
              & X-ray & 0.128  & 0.022 &10.2 \\
\noalign{\smallskip}
\hline
\end{tabular}
\end{table}

\subsection{The variability spectrum}
\label{VarspecSection}

\begin{figure}
\hspace*{2mm}\epsfig{angle=90,width=3.6in,file=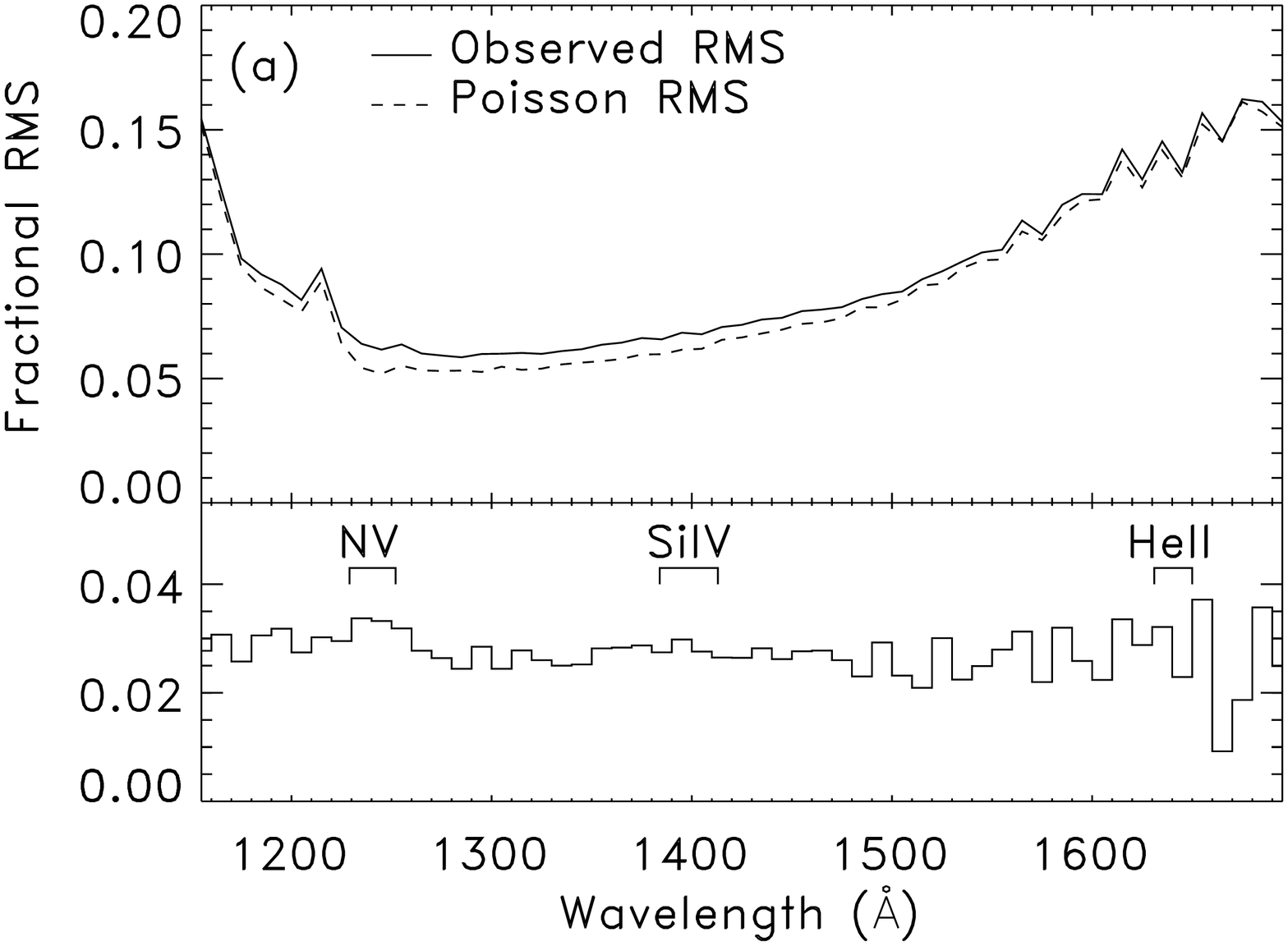}
\hspace*{2mm}\epsfig{angle=90,width=3.6in,file=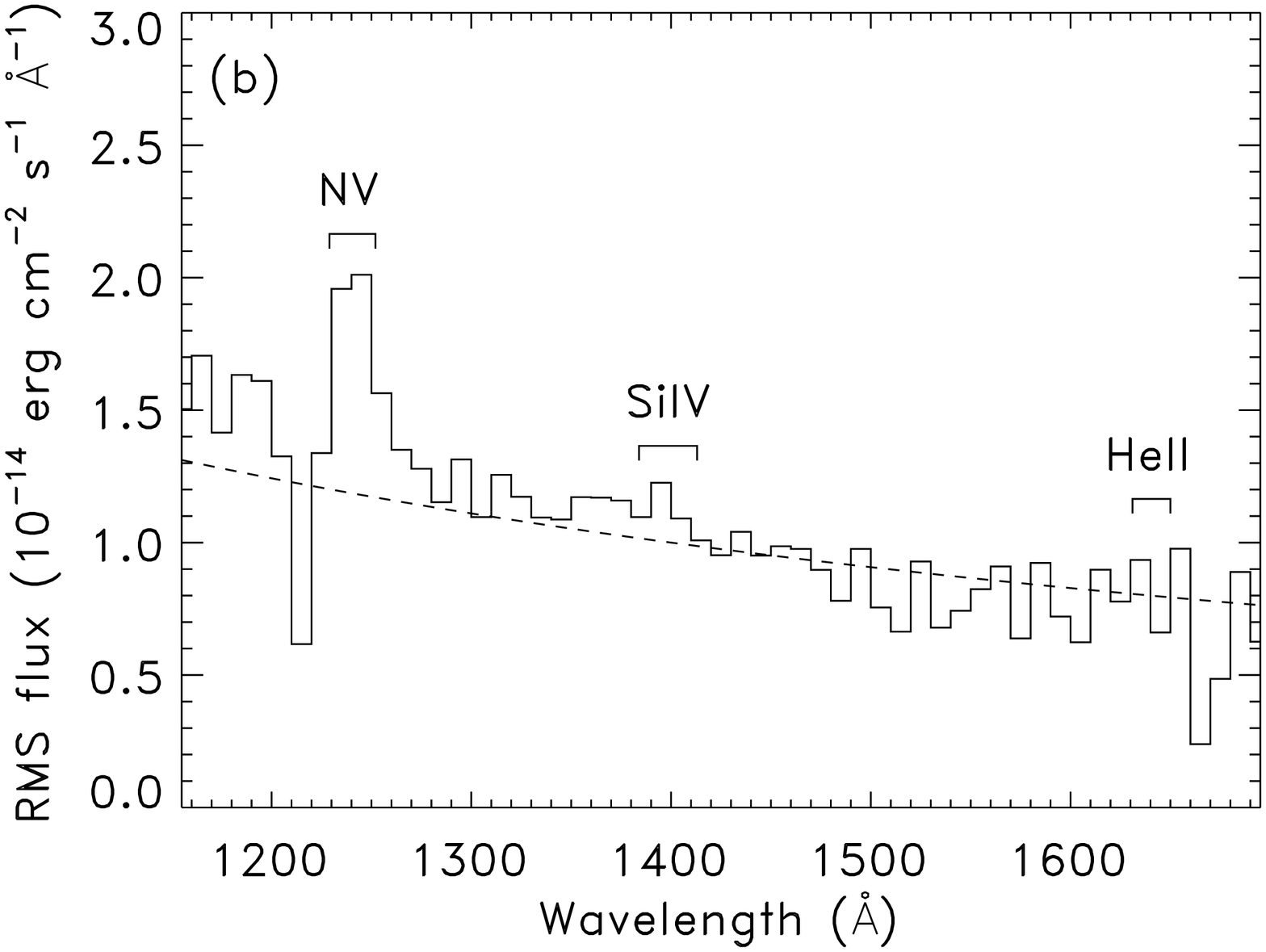}
\caption{ a) Fractional rms far-UV variability spectrum derived from
all E140M data.  The upper panel shows the measured rms as a function
of wavelength together with the expected rms if only Poisson noise
were present.  This rises at the ends of the wavelength coverage and
also at Ly$\alpha$, where interstellar absorption dramatically reduces
the count rate.  The lower panel shows the residual after removal of
the Poisson contribution.  The locations of the strongest emission
lines observed in the average spectrum are indicated; N\,{\sc v} may
be more variable than the continuum, but otherwise the fractional
variability appears almost gray.  b) The far-UV rms variability
spectrum, derived from all E140M data, converted to fluxes by
multiplication by the average dereddened spectrum.  The N\,{\sc v}
line is clearly present in the variable component.  The continuum
slope is poorly constrained without broader simultaneous coverage.
The dashed line shows a $F_{\nu}\propto\nu^{-0.59}$ spectrum,
consistent with the broad-band variability SED
(Fig.~\ref{RMSSEDFig}).}
\label{VarspecFig}
\end{figure}

As the far-UV spectrum of \target\ contains several emission lines,
principally the N\,{\sc v} resonance line but also weaker lines of
Si\,{\sc iv} and He\,{\sc ii} (Haswell et al.\ 2002), we examined
whether the far-UV variability is coming from the lines, the
continuum, or both.

We constructed rms spectra for each E140M visit using low spectral
resolution (10\,\AA) and a time-resolution of 5\,s.  The data quality
were insufficient for higher resolution.  The wavelength calibration
was done rather crudely by tracing the echelle orders and hence
assigning a wavelength to each pixel on the detector (and hence to
each detected photon) based on which order it is nearest to, but this
was adequate for the low spectral resolution achievable from the
time-resolved data.  The derived average far-UV variability spectrum
is shown in Fig.~\ref{VarspecFig}.  Significant variability above
Poisson noise is seen at all wavelengths with a sufficient count rate.
The fractional rms varies little with wavelength and the variability
spectrum has a similar shape to the average (see Haswell et al.\
2002).  No lines show {\em strong} excess variability, although
N\,{\sc v} does appear slightly more variable than the continuum.
N\,{\sc v} is clearly present in the variability spectrum, hence the
variability is not purely due to the continuum.  We will return to the
question of the origin of this line variability in
Section~\ref{LineEchoSection}.
%
%%%%%%%%%%%%%%%%%%%%%%%%%%%%%%%%%%%%%%%%%%%%%%%%%%%%%%%%%%%%%%%%%%%%%%%%%%%%%%%
%
\section{X-ray -- ultraviolet correlations}
\label{LagSection}
\subsection{Cross-correlations}
\label{CCFSection}

\begin{figure*}
\epsfig{angle=90,width=7in,file=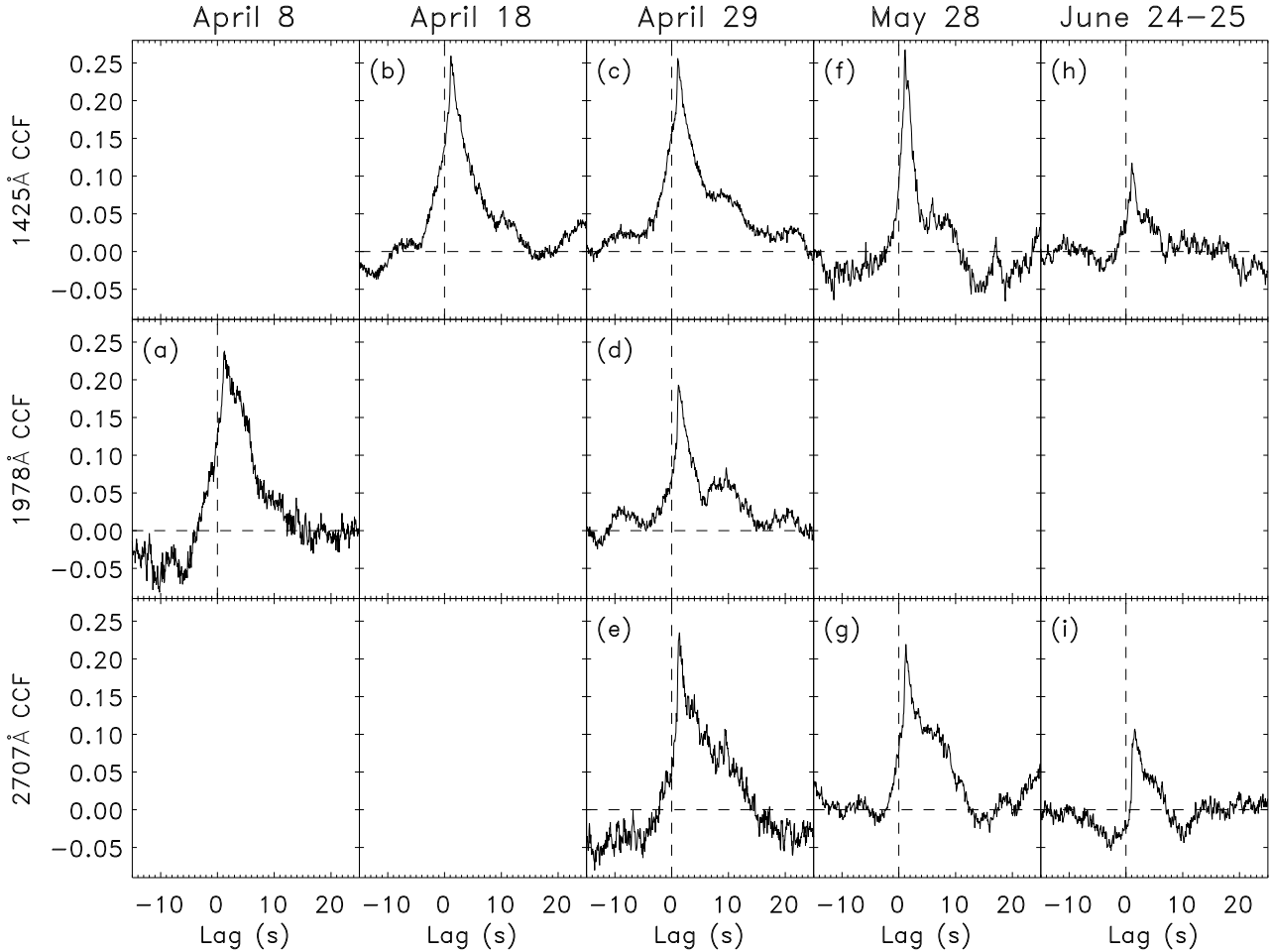}
\caption{X-ray and ultraviolet cross-correlation functions.  Results
from all simultaneous \HST/\RXTE\ observations are shown.  No
detrending or high-pass filtering has been done, hence the base level
is sometimes offset from zero.  The uncertainty in the \HST\ absolute
timing would allow an overall shift of all of the CCFs somewhat along
the lag axis.  The same will be true of all the CCFs and transfer
functions shown.}
\label{CCFFig}
\end{figure*}

For a crude analysis of simultaneous lightcurves we can construct
cross-correlation functions (CCFs), as was done for GRO\,J1655--40 by
Hynes et al.\ (1998).  This method has a weakness in that the CCF peak
is broadened by the low-frequency variability present, giving it a
width similar to the ACF of the lightcurves.  A better approach is to
properly model the echo light curve by convolving the driver
lightcurve with a transfer function.  The deconvolved transfer
function is considerably more informative than a CCF.  This analysis
will be performed in the next section, but a cross-correlation
analysis is a useful prelude and allows direct comparison with the
optical results of Kanbach et al.\ (2001).  It will also provide a
check of how well a linear reprocessing model (implicit in the
deconvolution) actually describes the data.

We constructed CCFs from each pair of 1/16\,s lightcurves.  We show
the central sections of the CCFs in Fig.~\ref{CCFFig}.  All show a
strong peak corresponding to UV lags of 1--2\,s (see
Table~\ref{LagTable}), although as discussed in
Section~\ref{HSTSection}, an overall offset of {\em all} the \HST\
times is possible, which would result in a shift of all the CCFs along
the lag axis.  There is also a notable asymmetry to all the CCFs, with
a fast rise and slower decay.  The form of the peak is thus similar to
that reported by Kanbach et al.\ (2001).  The lags detected are
wavelength dependent; on average a longer delay is seen at longer
wavelengths.  This result stands robustly despite the uncertainty in
the {\em absolute} timing of the UV lightcurves.  A linear fit to the
lags implies a rate of change of the delay of 0.16\,ms\,\AA$^{-1}$,
with a zero point offset of $\sim0.9$\,s; it is possible that the zero
point offset is entirely due to the timing uncertainty.  The CCF
morphology also appears wavelength dependent.  The asymmetry is more
pronounced at longer wavelengths; the far-UV peaks are almost
symmetrical.  Kanbach et al.\ (2001) have a `precognition dip' in
their CCF 2--5\,s before the peak.  This feature is not prominent in
our UV CCFs, although some suggest it and it may contribute to the
fast rise before the peak.

\begin{table}
\caption{Peak lags measured between X-ray and UV lightcurves using
CCF, MEM, and negative response deconvolution methods.  The UV
wavelength is given; all CCFs use the full PCA bandpass.  Note that
allowing a negative transfer function allows the peak to move
systematically to shorter lags compared to MEM.  The centroids remain
similar, however.  See text for details.  It must be remembered that
an overall offset of up to a few seconds could affect all the measured
lags, but that variations in the lag with time or wavelength are
robust.}
\label{LagTable}
\begin{tabular}{llrccc}
\hline
\noalign{\smallskip}
Date & $\lambda$ (\AA) & \multicolumn{1}{c}{Overlap} & \multicolumn{3}{c}{Lag (s)} \\
     &                 & \multicolumn{1}{c}{(s)} & CCF & MEM & Neg. \\
\noalign{\smallskip}
2000 April 8  & 1978 &  230 & 1.25 & 1.25 & 1.18 \\
2000 April 18 & 1425 & 1040 & 1.16 & 1.12 & 1.06 \\
2000 April 29 & 1978 &  820 & 1.26 & 1.19 & 1.13 \\
2000 April 29 & 2707 &  260 & 1.40 & 1.37 & 1.31 \\
2000 April 29 & 1425 & 1480 & 1.14 & 1.06 & 1.00 \\
2000 May 28   & 2707 &  750 & 1.31 & 1.25 & 1.19 \\
2000 May 28   & 1425 &  270 & 1.13 & 1.12 & 1.06 \\
2000 June 24  & 2707 &  700 & 1.57 & 1.56 & 1.51 \\
2000 June 25  & 1425 &  580 & 1.18 & 1.06 & 1.01 \\
\noalign{\smallskip}
\hline
\end{tabular}
\end{table}
\subsection{Maximum entropy echo mapping -- positive, linear responses}
\label{MEMSection}
A more powerful technique for analysing correlated variability is the
maximum entropy mapping technique (MEM; Horne, Welsh \& Peterson 1991;
Horne 1994), widely used in the analysis of AGN light curves.  While
the method was developed for the case where lags are due to light
echoes, this is not required, and the transfer function could equally
describe lags due to other effects.  The method does, however, assume
a linear, positive response (Blandford \& McKee 1982), i.e.\ that the
echo light curve is the convolution of the driver light curve with the
transfer function:
\begin{equation}
f_{\rm UV} = \Psi * f_{\rm X}
\end{equation}
where $\Psi\geq0$.  Since the near-UV lightcurves cannot be accounted
for with positive, linear responses (see Section~\ref{ACFSection})
this method cannot fully reproduce these data.  It can handle
low-amplitude non-linear responses by linearising them (Horne 1994),
but that is not applicable here where the X-ray variations are large.
We must, therefore, view the results of this analysis with caution.
 
We calculated transfer functions for all of our simultaneous
lightcurves.  In every case the response is dominated by a single,
very sharp peak with a lag of 1--2\,s.  The peak lags are summarised
in Table~\ref{LagTable} and are similar to those measured from the
CCF.  As discussed above, a MEM reconstruction is not strictly
applicable.  Consequently there will be problems with any data sets
for which the UV ACF is narrower than the X-ray one, and the method
should be most reliable for far-UV data.

\begin{figure}
\hspace*{-5mm}\epsfig{angle=90,width=3.6in,file=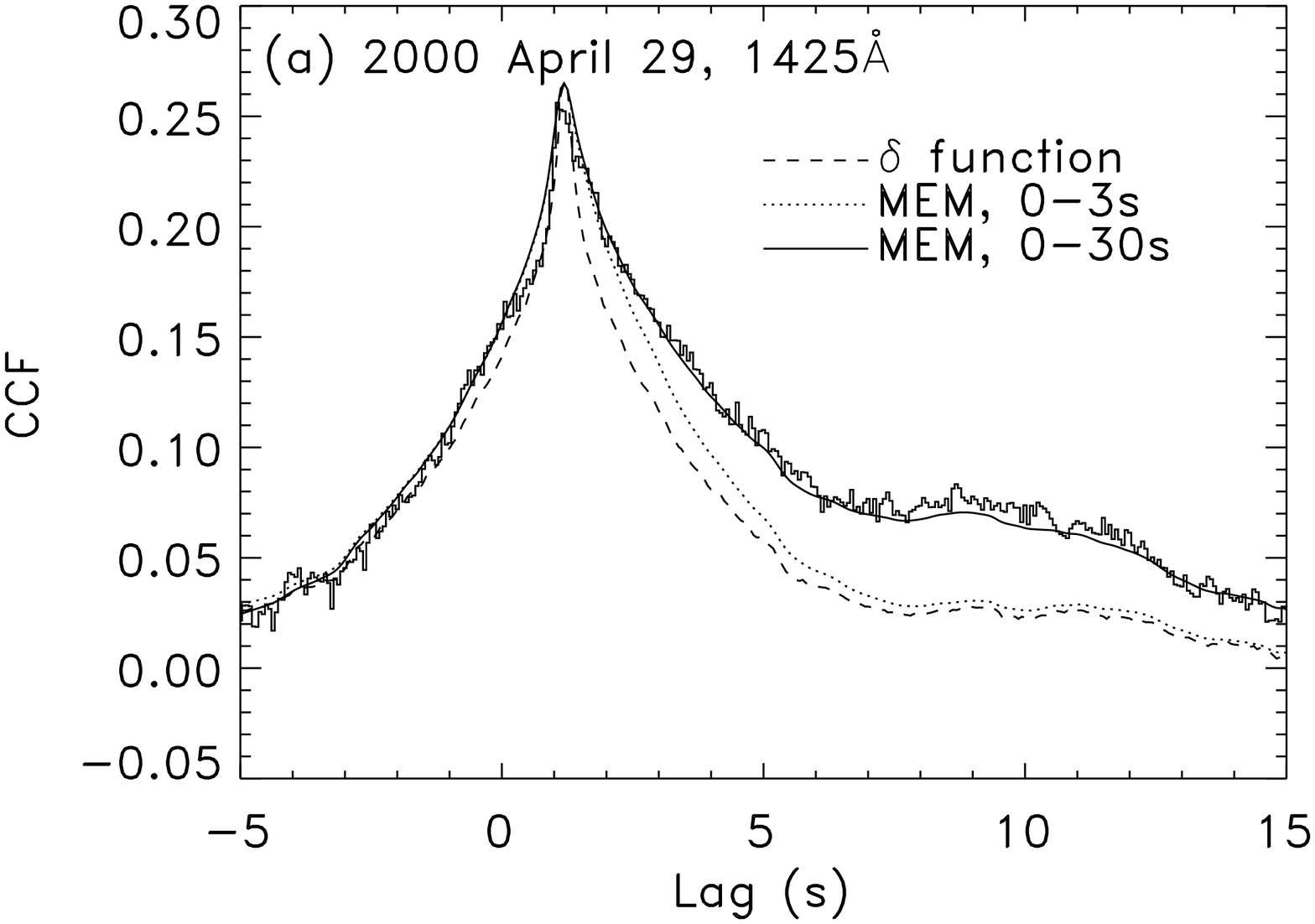}
\hspace*{-5mm}\epsfig{angle=90,width=3.6in,file=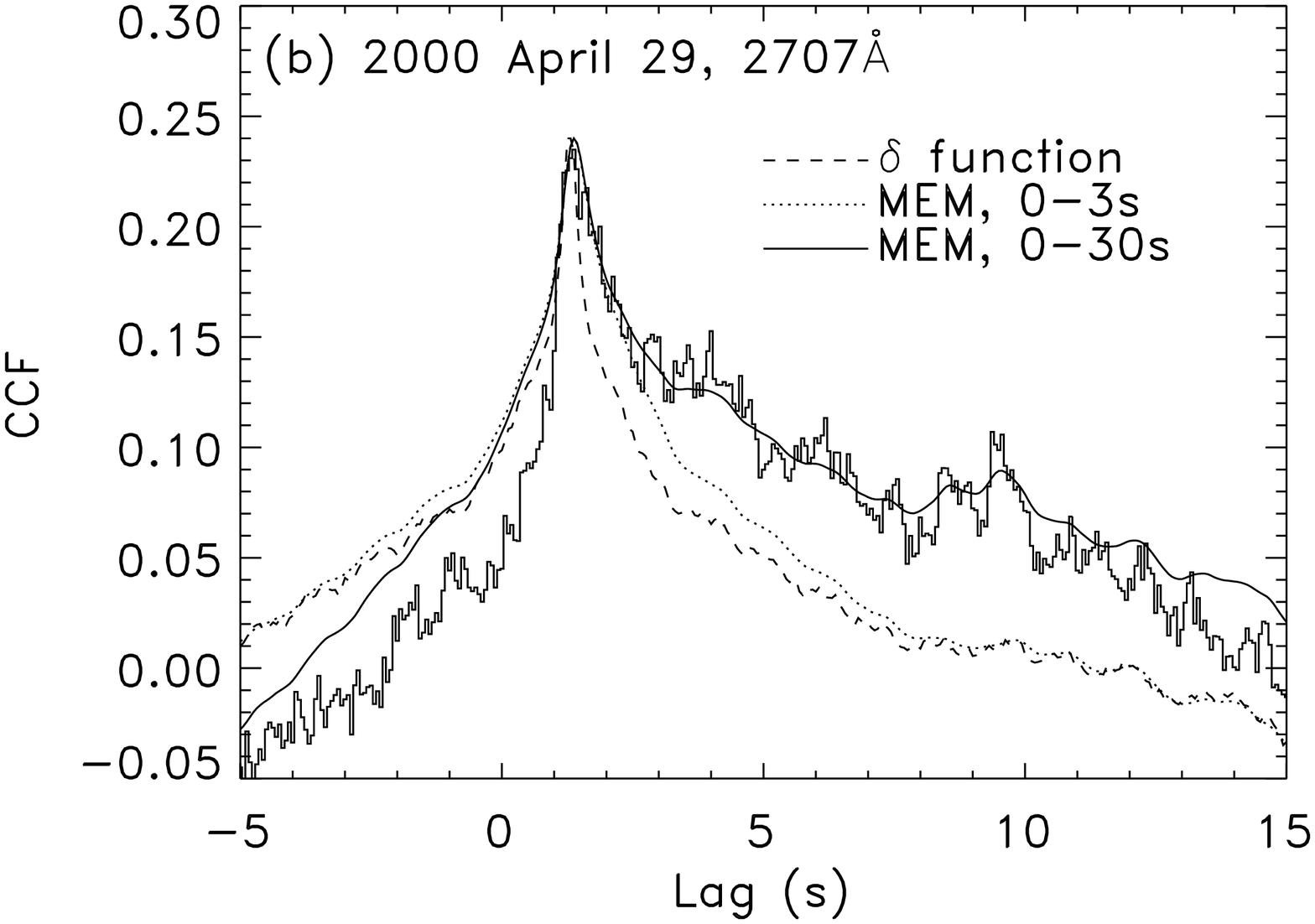}
\caption{CCFs from 2000 April 29 in far and near-UV.  The fits are the
X-ray ACF convolved with various transfer functions: a $\delta$
function chosen to reproduce the peak lag, the full (0--30\,s) MEM
transfer function and the same with only the 0--3\,s response used.}
\label{ModelCCFFig}
\end{figure}

The applicability of the method can be tested using the CCFs.  In a
simple linear reprocessing model, the ACF and CCF will be related by
the transfer function, $\Psi$: ${\rm CCF} = {\rm ACF_x} * \Psi$.  We
can test how successfully the model describes the data by comparing
the convolution of the X-ray ACF and the reconstructed transfer
function with the CCF.  If there is a good agreement then we can hope
that the transfer function recovered by MEM is a reasonable
approximation to the real one.  If the agreement is poor, then it is
likely that some of the assumptions involved are violated.  Indeed, we
already expect that this will be the case for those data sets with UV
ACFs which are narrower than the simultaneous X-ray ones.  We plot two
examples in Fig.~\ref{ModelCCFFig}.  These are far and near-UV CCFs
from the same visit, 2000 April 29.  A $\delta$ function response is
never adequate.  While the rise of the far-UV data can be reproduced
in this way, there is an extended tail which is not, indicating a
range of larger lags are present.  The MEM transfer function does
appear to provide an adequate fit for the far-UV data, suggesting that
the linear reprocessing model is applicable here.  For the near-UV
data, however, the model cannot account for the sharp rise in the
transfer function, so this transfer function should be considered
unreliable.  This bears out our assessment based on the ACFs: linear,
positive reprocessing cannot (entirely) account for the CCF when the
UV ACF is narrower than the X-ray one.  This rules out all of the
near-UV lightcurves and some of the far-UV ones for reliable MEM
deconvolution, at least of the main sharp peak.  The 2000 April 29
far-UV lightcurve can be well fitted by this method, and has a large
amount of overlapping data ($\sim1500$\,s), so is the best suited to
quantitative analysis of the transfer function.  The 2000 May 28
far-UV lightcurve is also reasonably fitted so can also be modelled,
although with less overlapping data ($\sim250$\,s) this is less well
constrained.  These are the only two lightcurves for which the UV ACF
is not narrower than the X-ray one and the only ones for which the CCF
can be well reproduced from the transfer function.  Hence it is
reasonable to believe that these can be treated in this way.

\begin{figure}
\hspace*{-5mm}\epsfig{angle=90,width=3.6in,file=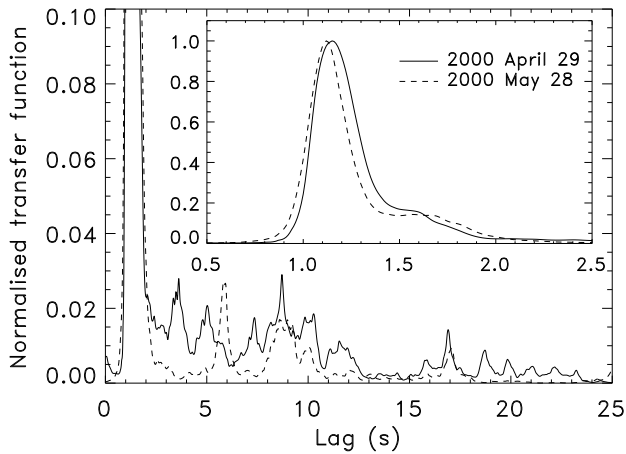}
\caption{MEM echo maps between X-rays and far-UV for the data sets for
a which linear reprocessing model appears applicable.  Both are in the
far-UV band.  The main panel shows a large time range to illustrate
the extended response to $\sim12$\,s, the inset shows the main peak.
As for the CCFs, an offset along the lag axis is possible.}
\label{MEMFig}
\end{figure}

The useful far-UV MEM transfer functions are shown in
Fig.~\ref{MEMFig}.  They are dominated by a very sharp response around
1.2\,s, corresponding to the peak of the CCF.  The structure of this
peak is very similar for the two good far-UV transfer functions, with
a main peak and a tail extending to $\sim2$\,s.  There also appears to
be an extended response at up to 12\,s or more.  Since the MEM
algorithm is free to add an additional response anywhere in the
transfer function if it significantly improves the $\chi^2$ of the
fit, some spurious noise features can be expected at a low level.  To
test if the features observed are real or not we show in
Fig.~\ref{ModelCCFFig} the effect of only using the well determined
peak of the MEM transfer function (0--3\,s) in attempting to reproduce
the CCF.  When this is done the peak of the CCF is reproduced, but the
extended response at positive lags is not.  Since this is reproduced
with the 0--30\,s transfer function, it must be coming from the low
level response seen at lags $\ga3$\,s.  This extension of the CCF to
larger lags is a ubiquitous feature of the CCFs so is seems that there
must be a real UV response at lags $>3$\,s, although it cannot
be recovered in detail from these data.

To test how well the MEM method can recover transfer functions from
the data used we performed a number of simulations based on the data
from 2000 April 29 as suggested by Horne (1994).  We take the MEM
reconstructed X-ray lightcurve as a starting point and convolve it
with a number of test transfer functions to model the UV lightcurve.
We add an additional background contribution such that the fractional
rms variability of the modelled UV lightcurve matches that observed,
3.6\,percent.  We then normalise to match the observed countrate, add
noise to both lightcurves assuming Poisson statistics, and attempt to
reconstruct the original transfer function using MEM deconvolution.
The simplest test is to assume that the UV lightcurve is simply equal
to the X-ray one, plus a non-varying background.  This corresponds to
assuming a $\delta$-function response.  The transfer function
recovered in this case was a quasi-Gaussian shape, with width
comparable to the main peak of the transfer functions recovered from
the data.  This is known as the impulse test (Horne 1994) and
indicates the resolution achievable from the data.  In this case it
suggests that the main peak of our reconstructed transfer functions is
not resolved, and could actually be much sharper than shown in
Fig.~\ref{MEMFig}.

Next we used a Gaussian transfer function with $\sigma=1$\,s (i.e.\
FWHM$\sim2.35$\,s).  The MEM deconvolution reconstructed this very
well, indicating that a broad transfer function could be recovered
from these data if present.  Thus we are not `over-resolving' the
peak; it really is narrow.  We also performed tests with more
realistic model transfer functions; these will be described in
Section~\ref{DiscussionSection}.

A disadvantage of MEM echo mapping is that it does not produce a
unique solution, but rather a family of solutions, all of which fit
the data ($\chi_R^2=1.0$).  These solutions differ in the relative
weights assigned to the entropy of the reconstructed X-ray lightcurve
and the reconstructed transfer function.  We tested reconstructions
using a range of weights.  All reconstructed transfer functions are
asymmetric, but the form of the asymmetry, whether two distinct peaks
or one with a fast rise and slow decay, is not well determined.  The
width of the transfer function is also poorly constrained, although
the broadest transfer function obtained has a FWHM of only 0.4\,s and
no reconstruction produces a strong response (relative to the main
peak) outside the range 0.8--2.0\,s.

\subsection{Non-linear responses}
\label{NonLinearSection}
One way to produce an optical/UV ACF narrower than that seen in X-rays
is if the optical/UV responds non-linearly, for example if $f_{\rm
opt} \propto f_{\rm X}^2$.  To test this possibility we have
constructed simulated datasets as described earlier in which the
response was assumed to be a $\delta$ function in time, but in which
the strength of the response varied as some power, $n$ of the input
flux: $f_{\rm opt} \propto f_{\rm X}^n$.  From simulated lightcurves
we then constructed X-ray and optical/UV ACFs and compared them with
our UV observations.  We also compare with the optical ACF of Kanbach
et al.\ (2001); since we are using X-ray lightcurves obtained earlier
in the outburst, there will not be an exact correspondence with this,
but the general trend can be compared.  Because we are using the MEM
reconstructed lightcurves (to remove the noise) the shortest timescale
variability has been smoothed out and so the peaks are rounder than
actually observed.  Taking this into account, this non-linear response
does provide a reasonable fit to other aspects of the ACF
phenomenology.  The wings of the optical response can be greatly
suppressed, although a relatively large power, $n\ga4$ would be needed
to match the optical data.  Also, since the ACF of Kanbach et al.\
(2001) actually drops below 0 for lags beyond 1\,s an exact match
cannot be made; this approach always produces an optical ACF which is
positive if the X-ray one is.  There is also no way to produce a
precognition dip in this way.  Therefore a non-linear response cannot
provide a full description of the behaviour of \target.

\subsection{Linear, negative responses}
\label{NegativeSection}

If non-linear responses cannot entirely describe the observed
phenomenology then we must turn to negative responses.  A negative
response has been suggested, in the form of a precognition dip, by
Kanbach et al.\ (2001).  We therefore tried creating simulated
lightcurves using a transfer function which includes such a dip.  This
proved extremely successful, as including this produces both a faster
rise in the CCF and a narrower UV ACF.  With a strong enough dip, the
ACF can drop below zero as seen by Kanbach et al.\ (2001).

To further explore this possibility we developed a method to
reconstruct transfer functions without the explicit positivity
constraint of MEM.  We formulate the problem rather differently to the
MEM method.  We do not convolve the X-ray lightcurve with the trial
transfer function to obtain the UV lightcurve; we instead convolve the
X-ray ACF with the transfer function to obtain the CCF.  This method
has several advantages.  Because we have a lot of data, but of
relatively low quality, the average X-ray ACF is of high quality, but
the lightcurves are relatively poor.  Consequently to use the
lightcurves directly it is necessary not only to fit for the transfer
function but also to recover the (noise-free) X-ray lightcurve.  Since
the UV lightcurve then depends on the convolution of two fitted
functions, the problem is inherently non-linear.  We can, however,
assume the X-ray ACF to be noise-free, so that we then must only solve
a linear problem to obtain the transfer function.  This linearity,
combined with the reduced number of points to be fitted makes the
solution easier.

Removing the positivity constraint does give more freedom as solutions
are possible which have zero mean response, but oscillate as necessary
to fit the noise.  Our formulation should reduce these noise
oscillations (compared to fitting lightcurves), but it is also
necessary to employ a maximum entropy regularisation term.  With each
iteration we calculate a default image which is a Gaussian-blurred
version of the current transfer function.  The reconstruction is
driven by a combination of trying to obtain a good fit to the measured
CCF and trying to obtain a transfer function which is relatively
invariant to this blurring, i.e.\ trying to obtain a smooth transfer
function.

\begin{figure}
\hspace*{-5mm}\epsfig{angle=90,width=3.6in,file=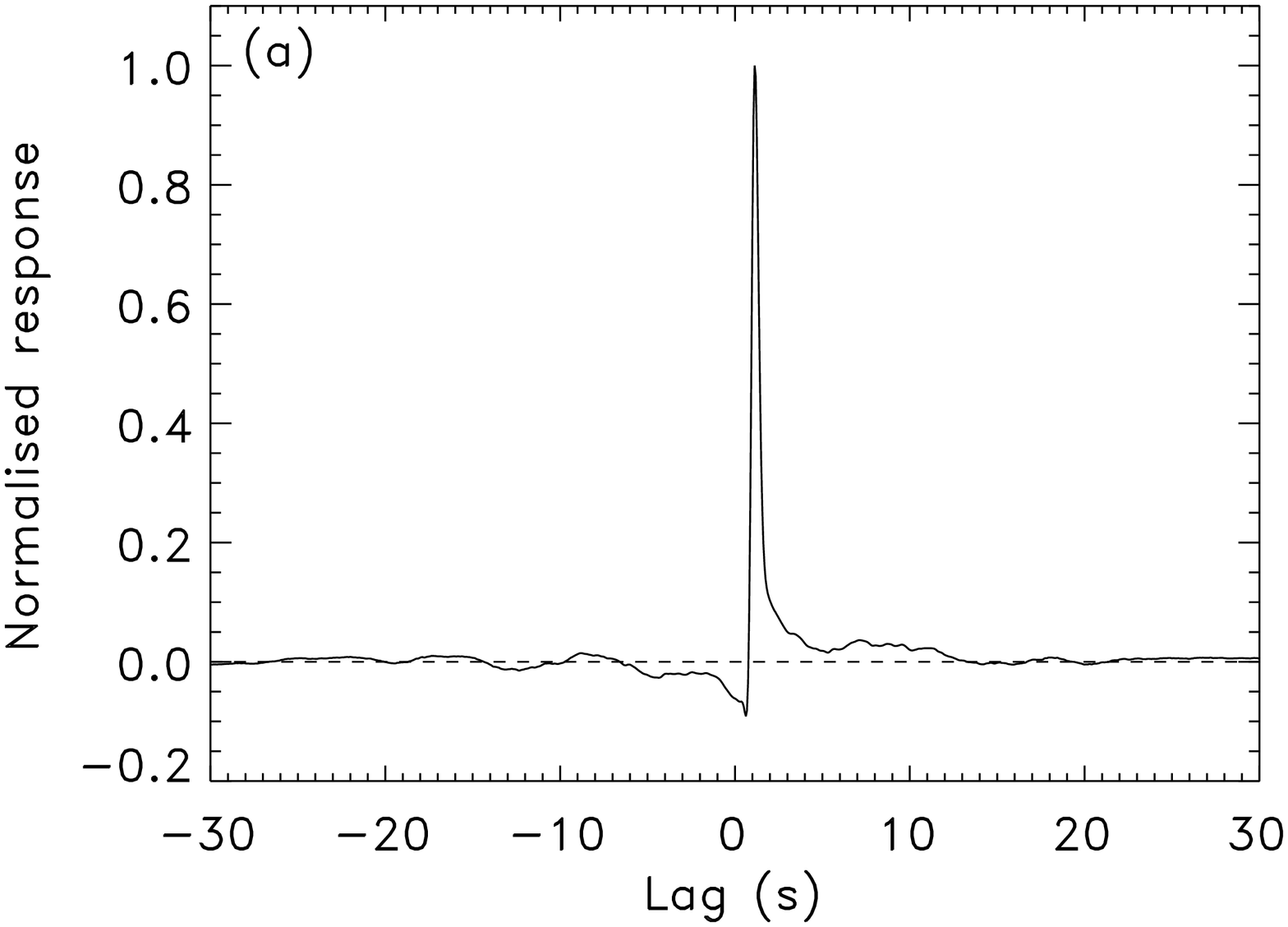}
\hspace*{-5mm}\epsfig{angle=90,width=3.6in,file=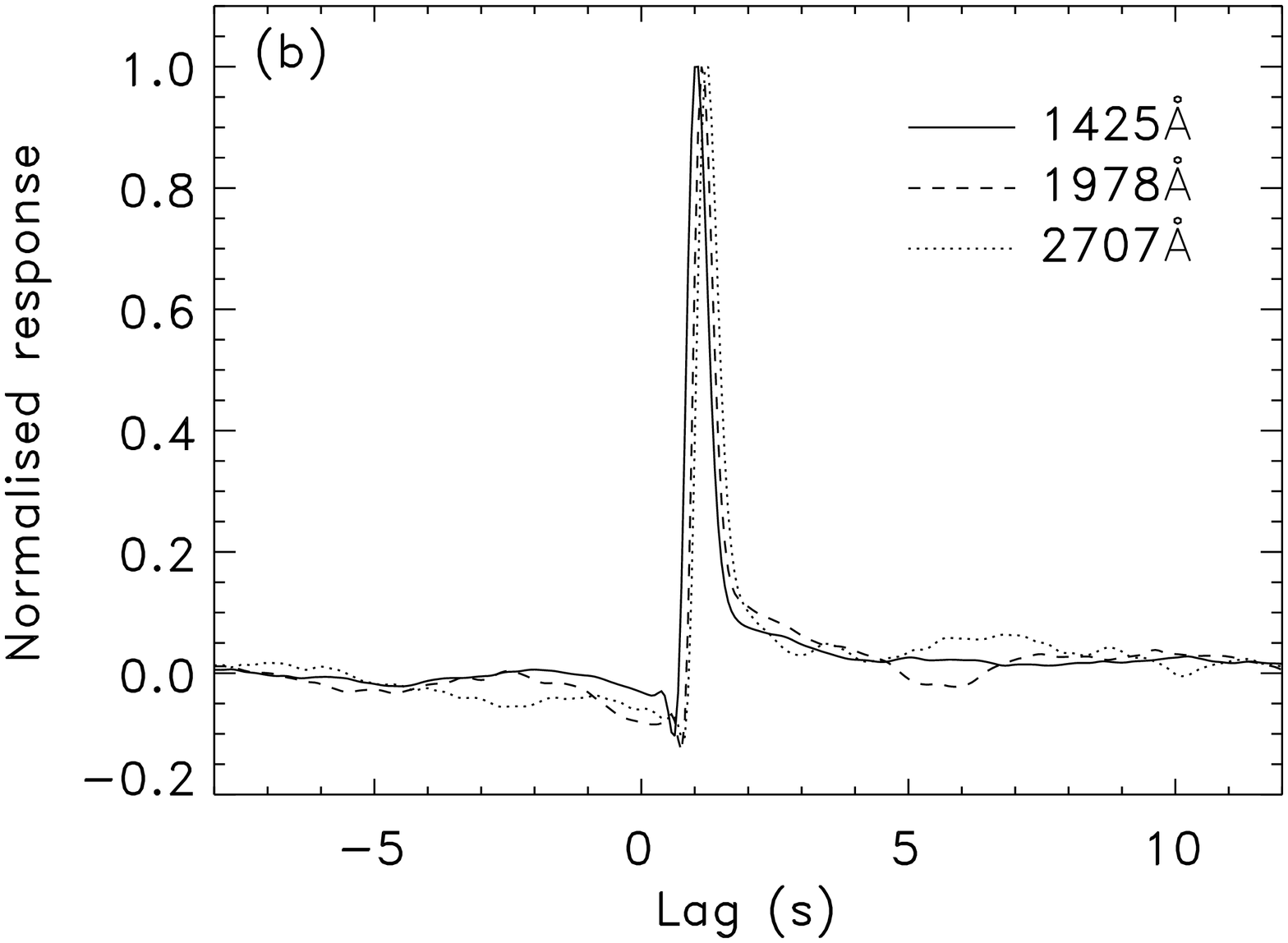}
\hspace*{-5mm}\epsfig{angle=90,width=3.6in,file=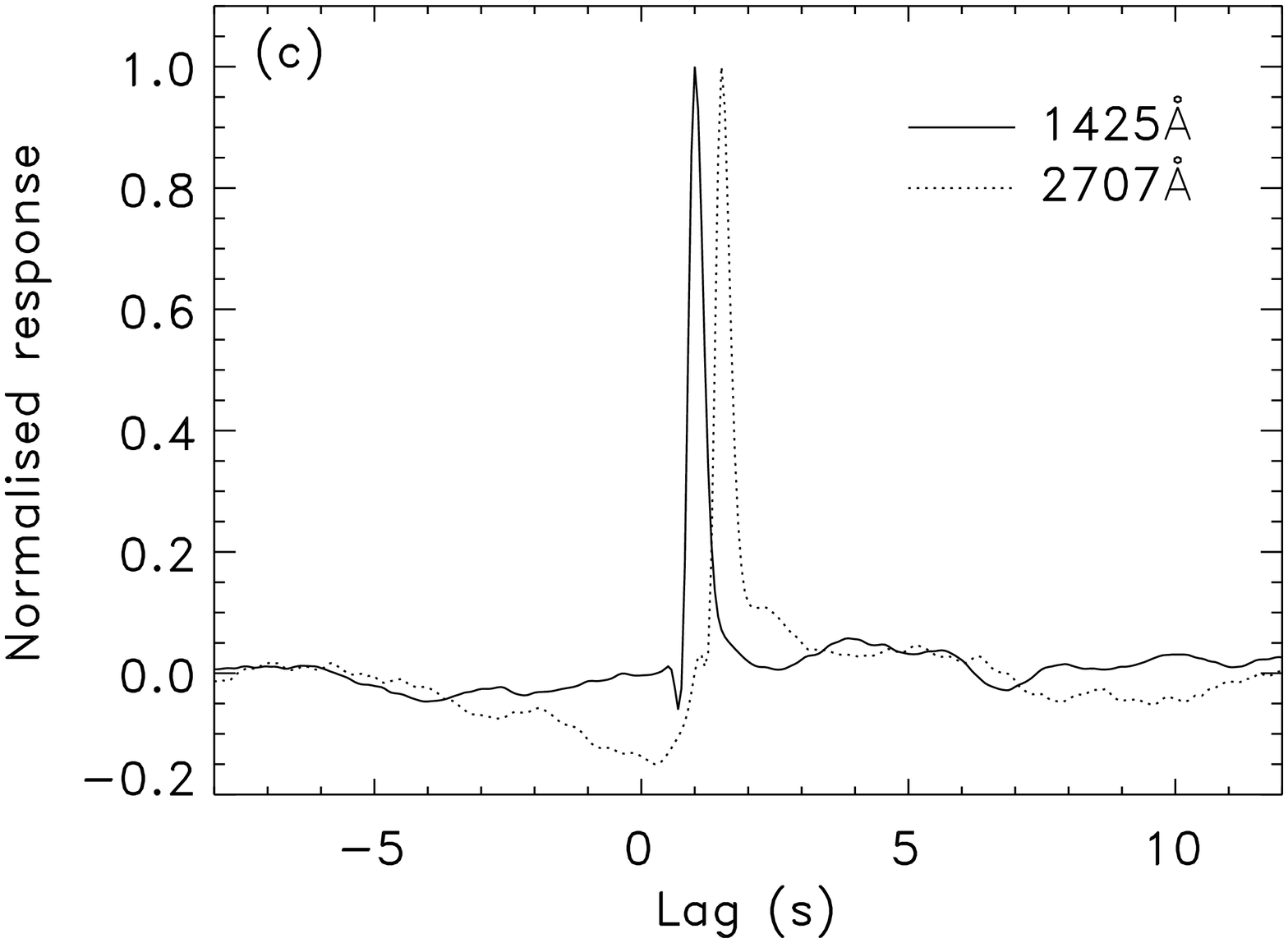}
\caption{Transfer functions reconstructed allowing negative responses.
a) Exposure time weighted average response for all wavelengths from
the first four visits.  b) The same but subdivided by wavelength and
with a magnified time axis.  c) Transfer functions from the last
visit.  The notches seen immediately before the peak in some cases are
likely to be artifacts of the reconstruction.  As for CCFs and MEM
transfer functions, an offset along the lag axis is possible.}
\label{NegFitFig}
\end{figure}

With this method we find broadly similar results to those obtained
with MEM.  Some of the reconstructed transfer functions are shown in
Fig.~\ref{NegFitFig}.  All are dominated by a strong peak around
1--2\,s lag, with a longer lag at longer wavelengths.  There is also
an extended response up to 10\,s.  There is however, an additional
component: a dip before the main peak.  As expected, this is most
prominent at the longest wavelengths where the ACFs are narrowest.
The last visit near-UV transfer function appears rather different to
earlier ones.  The lag is larger and the precognition dip is much more
pronounced.  This suggests that the transfer function is evolving at
the end of the outburst, which may also contribute to the differences
between our UV transfer functions and the optical results of Kanbach
et al.\ (2001).  Most of the reconstructed transfer functions show
small, sharp, negative notches immediately before the peak.  This
appears to arise from the difficulty of fitting the rapid rise in the
CCFs and is likely not a real fast dip.  It may simply arise from the
smoothing inherent in the regularisation, but could also occur if the
rise is somewhat faster than linear.

In conclusion, provided negative responses are allowed, we can
reproduce the properties of the UV variability using a model in which
the X-ray variations are convolved with a transfer function.  This
transfer function appears to have three repeatable components.
Firstly there is a weak UV dip occuring up to $\sim10$\,s before an
X-ray event.  This appears most prominent at longer wavelengths and
appears to be the counterpart to the much stronger dip seen in the
optical data of Kanbach et al.\ (2001).  Secondly a strong, sharp UV
peak occurs about 1.0--1.5\,s after the X-ray peak.  This occurs later
at longer wavelengths.  Thirdly there appears to be a more extended
response, continuuing for 10\,s or more after an X-ray event.  We
should caution, however, that all of the techniques we have used are
based on statistical analysis of the lightcurves, and so many events
contribute.  The transfer function may thus be an average of a range
of different behaviours; this is suggested by a comparison of the
lightcurves in Fig.~\ref{RapidLCFig}; there is certainly not a
one-to-one correspondence between X-ray and UV lightcurves and X-ray
events may produce little or no response and vice-versa.  The
non-unity coherence at high frequencies also indicates that
uncorrelated variability is present.  Consequently, all three
components of the UV `response' may not always be present, and broad
features, e.g.\ the dip, may actually represent sharp features at a
range of lags.  The sharpness of the peak, however, does indicate a
relatively constant delay, which is borne out by the consistency of
lags measured in the same bandpass at different epochs.  It is also
possible that the different components of the response have different
origins, for example thermal reprocessing in the disc and/or UV
synchrotron variability correlated with the X-ray emission could both
be present.

\subsection{Cross spectral analysis}
\label{CrossSpecSection}

\begin{figure}
\epsfig{width=3.2in,file=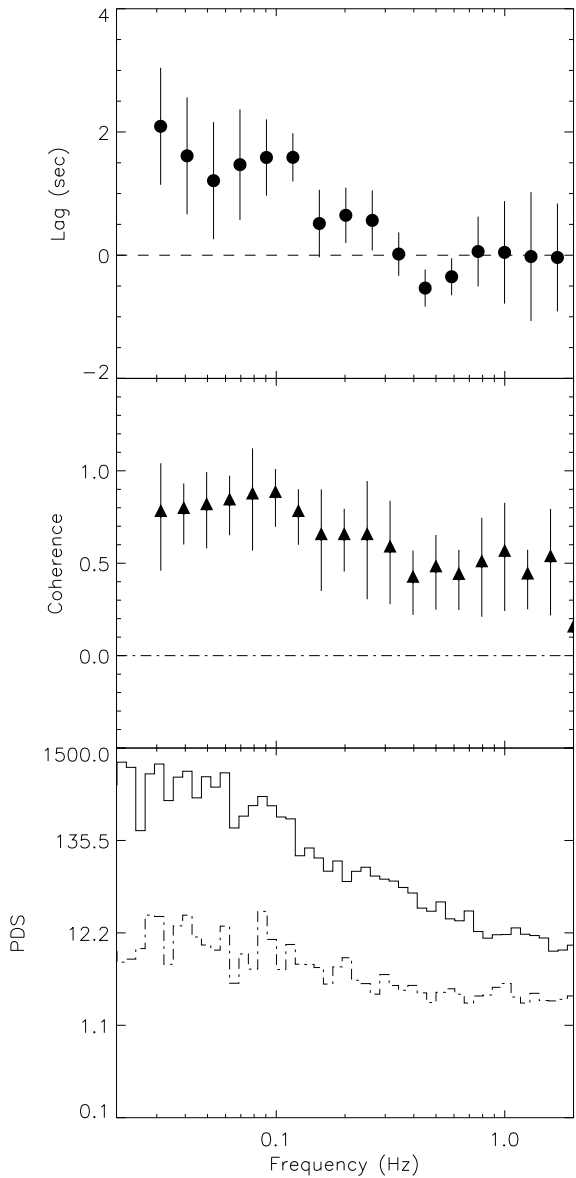}
\caption{The upper two panels show the lag spectrum and the
coherence derived from the complex cross-correlation analysis
described in Section~\ref{CrossSpecSection}. In the bottom panel the
PSD for the X-ray (upper) and far-UV (lower) are shown. The data shown
are from 2000 April 18. The $\sim0.1$\,Hz QPO feature is evident,
albeit marginally so. Note that the PSDs presented here are different
from the more rigorously determined cases of Figs.~\ref{PSDFig} and
\ref{MWPSDFig}, in that they are derived from only the HST/RXTE
overlap regions of one epoch.  Leahy et al.\ (1983) normalisation is
used.  Note that the lag spectrum, although limited by statistical
uncertainty, seems to peak in the $\sim0.1$\,Hz vicinity of the QPO,
and it appears that the $\sim 1$\,s lag seen in the cross-correlation
analysis emanates predominantly from $\sim0.05-0.3$\,Hz region. Of
notable interest is the possible negative lag near about
0.5--6\,Hz. This may be associated with the `precognition dip' noted
by Kanbach et al.\ (2000). It is curious that the coherence function
seems to decrease at frequencies coincident with the negative lag, as
the dips do appear correlated with X-ray flares.  This occurs close to
the range that will be affected by the likely timing error, though,
and as noted, the statistics are poor and the degree of coherence may
still be non-negligible.}
\label{CrossSpecFig}
\end{figure}

To further investigate the relative UV and X-ray variability
characteristics, and in particular the frequency dependence of the
time correlation between the two signals, we have computed the
(complex) cross-correlation function and the coherence function
following the procedure detailed in Bendat \& Piersol (1986); also see
Vaughan \& Nowak (1997). The two time series, prepared as described in
Sections~\ref{HSTSection} and \ref{XTESection}, were separated by
observation epoch, and then further broken into segments of lengths
(in bins) from 4096 to 32768. For the UV series, $o(t)$ and the X-ray
series, $x(t)$, one computes the Fourier transforms $O(f)$ and $X(f)$.
The cross spectrum is then $\left|O^{*}(f)X(f)\right|$, which can be
expressed as a time delay by dividing by $2\pi f$. This then allows us
to explore directly the frequency dependence of the overall
optical-to-X-ray time lags (see e.g.\ Kazanas \& Hua 1999).  There is
a caveat to this analysis.  The CCF and deconvolution approaches to
examining correlated variability are affected straightforwardly by the
uncertainty in the \HST\ timing; an error on the time translates
simply into an offset of the CCF.  The effect on the cross-spectrum is
less obvious however, since there is an ambiguity in the conversion of
phase-lags back into time-lags.  Since the error could be as large as
1\,s, at frequencies above 1\,Hz there will be complete uncertainty
about the phase lag.  Consequently, only values for frequencies
$\la1$\,Hz\ should be considered.

From the complex cross-spectrum, and the Fourier power density spectra
of the two signals, one can compute the coherence function.  This
provides a frequency-dependent measure of the degree of linear
correlation between the UV and X-ray signals, which when considered in
conjunction with other observable properties of the binary system
provide insight into the underlying physical processes.  The
normalisation, and the statistical uncertainty are then computed
following Bendat \& Piersol (1986).
               
We find upon examining the cross spectral lags for various
sub-segments of the data at various epochs that the 1--2\,s lags
identified in the cross-correlation analysis
(Section~\ref{CCFSection}; Fig.~\ref{CCFFig}) are evident in the
cross-spectrum at low frequencies, up to a few tenths of a Hz. This is
the frequency domain region containing the $\sim0.1$\,Hz QPO feature
described.  The coherence function is also highest at the lowest
frequencies, and remains significant up to $\sim0.5 Hz$.

More interestingly, there does appear to be evidence, albeit
tentative, for a negative lag feature at $\sim0.5$\,Hz. Although this
is marginal in a number of the individual lag-spectra that we
examined, it seems to be consistently present in all cases. An example
is illustrated in Fig.~\ref{CrossSpecFig}, where we have plotted the
lag, coherence, and power spectra for 2000 April 18. We tentatively
suggest that this negative lag may be associated with the
`precognition dip' identified by Kanbach et al.\ (2001); see also
Spruit \& Kanbach (2002). These authors speculate on possible
scenarios for its cause; the simplest perhaps being some sort of surge
in the in the accretion flow preceding an X-ray flare.  However, while
this is may be plausible in the context of reprocessing scenarios it
is more difficult to envision for synchrotron models.

Reprocessing-dominated optical/UV emission should have a high degree
of coherence with respect to X-rays (coherence function approaching
unity), whereas optical/UV variability resulting from variable local
viscous heating should have a coherence function (with respect to
X-rays) which tends towards zero. Other mechanisms, such as Compton
scattering could also lead to X-ray/UV coherence, although in this
case the sense of the lag is wrong, as high energies should lag the
low energies if X-rays arise from inverse-Compton scattered optical/UV
photons. A synchrotron source could also plausibly lead to a coherent
variability, in which case the time lag may be related to the electron
synchrotron lifetimes (and thus the magnetic field strength).  If the
higher energies were due (wholly or partially) to self-Comptonization
effects, again the sense of the $\sim0.5$\,Hz lag, assuming it is
real, is wrong.  It is also possible that a single synchrotron flare
spanning the optical to X-ray range (c.f.\ Markoff, Falcke \& Fender
2001) leads to the apparent low-frequency coherence; in this case the
lags are in the right sense if they represent the flare becoming
optically thin at progressively longer wavelengths (and always thin in
X-rays).
%
%%%%%%%%%%%%%%%%%%%%%%%%%%%%%%%%%%%%%%%%%%%%%%%%%%%%%%%%%%%%%%%%%%%%%%%%%%%%%%%
%
\section{Is there a signal from thermal reprocessing?}
\label{ReprocessingSection}

Correlated X-ray and optical/UV variability in LMXBs is usually
attributed to thermal reprocessing in the accretion disc and/or
secondary star.  Disc emission should produce a range of lags
dependent on the disc size; emission from the companion star will
produce a sharper response moving with orbital phase.  For \target, we
can predict lags assuming parameters from the literature: an orbital
period of 0.1699\,days (Zurita et al.\ 2002), a mass function
$(6.1\pm0.3)$\,M$_{\odot}$ (Wagner et al.\ 2001), mass ratio
$(0.037\pm0.008)$ (Orosz 2001), and inclination $71-82^{\circ}$
(Zurita et al.\ 2002).  A simple Monte-Carlo simulation assuming these
parameters and the quoted uncertainties implies a probable binary
separation of $a=(5.85\pm0.13)$\,ls. The disc radius is obviously
uncertain, but the presence of superhumps in outburst suggests it is
larger than the radius of the 3:2 resonance (Whitehurst \& King 1991),
and it should be smaller than the tidal truncation radius.  This
translates into a range of disc radii of 2.7\,ls$ < R_{\rm disc}
<$3.5\,ls.  For a flat, centrally illuminated disc, disc emission is
expected to extend to $R_{\rm disc} (1+\sin i)$, or a range of lags of
5.4--6.9\,ls.  The companion star will move within the range $a(1\pm
\sin i) \sim (0.1$--11.7)\,ls.
\subsection{Reprocessing on the companion star}

\begin{figure}
\hspace*{-5mm}\epsfig{width=3.6in,file=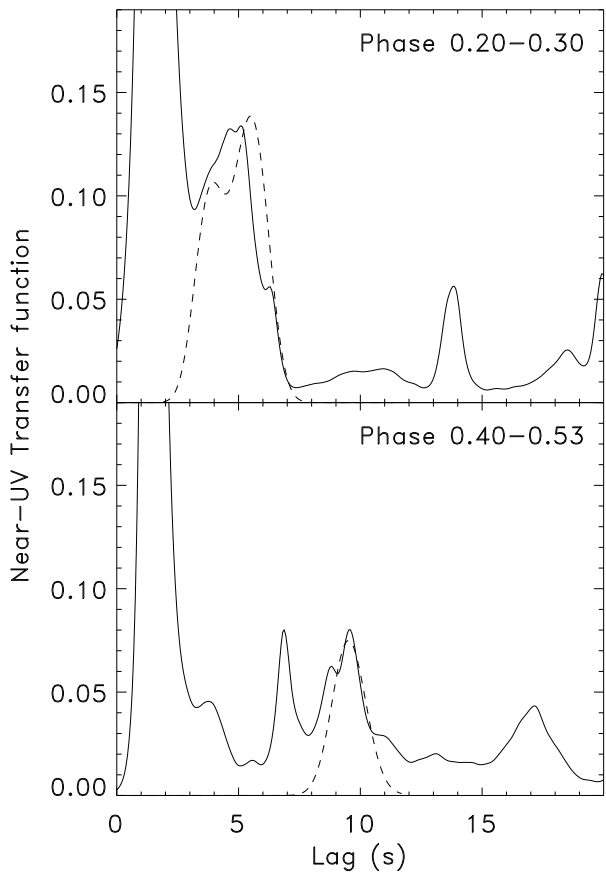}
\caption{MEM near-UV transfer functions averaged by orbital phase.
The solid line shows the average reconstruction for each phase range.
The dashed lines show synthetic companion star transfer functions
averaged over the phases observed, smoothed, and arbitarily scaled.
The data are consistent with a weak bump moving with the companion
star, but given the other structure present in the transfer functions,
this is not a conclusive detection.  Note that in the upper panel, two
non-overlapping phase ranges have been averaged, so the model transfer
function contains two features with a small dip between them.}
\label{CompanionFig}
\end{figure}

We have calculated model transfer functions as a function of orbital
phase using the code described by O'Brien et al.\ (2002) and the
parameters discussed above, and the results bear out the crude
estimates presented above.  The companion star should trace out a
large range of lags, and its expected motion is clearly inconsistent
with the stability of the main peak.  Some of the extended response we
observe, however, could originate in this way.  To test this we
calculated the phases of our overlapping segments using the ephemeris
of Zurita et al.\ (2002).  Our phase coverage is obviously extremely
patchy, and comprised of observations at different epochs and
wavelengths.  Unfortunately, all of our far-UV observations spanned
phases 0.69--1.09, whereas our mid- and near-UV ones cover two blocks,
0.20--0.30 and 0.40--0.53.  We disregard the far-UV data, as it will
be impossible to disentangle phase-dependence from wavelength
dependence, and in any case the companion star echo will be at very
short lags for these phases.  The two mid- and near-UV blocks,
however, should have the companion star echo significantly separated
from the main peak, and each includes both 1978\,\AA\ and 2707\,\AA\
observations.  We calculated an average of UV MEM transfer functions
for each phase block.  Similar, but less clear results were obtained
with a deconvolution allowing negative responses.  We show these
average transfer functions in Fig.~\ref{CompanionFig} together with
the model transfer function for the companion star only, averaged over
the phases observed.  Both phase blocks do appear to exhibit excess
response at the predicted lags.  We {\em do not} claim that this is a
conclusive detection of the companion star echo, but the results would
be consistent with and suggestive of this interpretation.  The
companion star clearly cannot explain the main peak, however.

\subsection{Reprocessing by the disc}

\begin{figure}
\hspace*{-5mm}\epsfig{width=3.6in,file=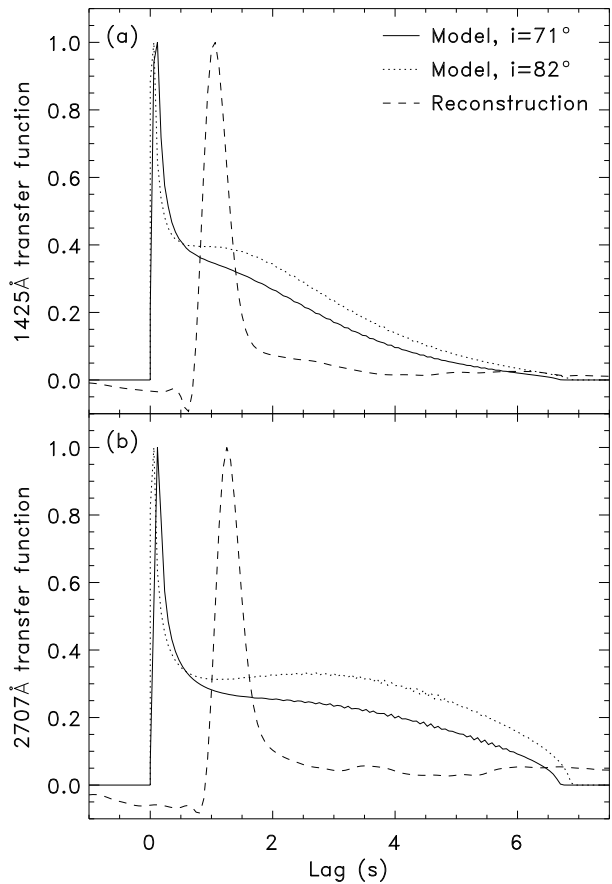}
\caption{Model disc transfer function for \target\ compared with
those reconstructed in Section~\ref{NegativeSection}.  a) and b) show
far- and near-UV respectively.  As well as the inclinations shown, a
range of disc paramters were considered.  In all cases the main spike
was close to zero lag and there is a strong response up to 7\,s from
the outer disc.  These are clearly not consistent with the
characteristics of the data.}
\label{KSOFig}
\end{figure}

We next moved on to consider if the main peak could be associated with
reprocessing from the disc.  A very low level of reprocessing is
already suggested by Miller et al.\ (2002) who place very tight upper
limits on the fraction of X-rays reflected by the disc
($\la0.5$\,percent).  To address this issue with our data we
calculated transfer functions for the disc only using a range of disc
thicknesses ($H/R=0.01-0.1$), and curvatures.  For details of how the
disc is modelled see O'Brien et al.\ (2002).  Transfer functions were
calculated for the \HST\ E140M/1425 and E230M/2707 bandpasses to test
wavelength dependence.  Selected results are shown in
Fig.~\ref{KSOFig}.

For all parameters considered, the main peak occurs at very short lags
and there is a pronounced extended response to $\sim7$\,s which is not
present in the data.  The peak occurs at very short lags because the
light-travel delay for material on the near-side of the disc is very
short for high-inclination systems.  This is true even for the
$i=71^{\circ}$, $H/R=0.01$ case; larger inclinations or disc
thicknesses push the peak closer to zero.  Because this is the
dominant factor in the peak delay, there is little wavelength
dependence, and the inferred $\sim0.2$\,s delay from 1425\.\AA\ to
2707\,\AA\ (c.f.\ Fig.~\ref{NegFitFig}) does not occur in the models;
hence even with an offset in the absolute times, the data cannot be
reconciled with the models at multiple wavelengths.  If the central
disc is missing, as suggested by the spectral energy distribution
(Hynes et al.\ 2000; McClintock et al.\ 2001b; Esin et al.\ 2001;
Chaty et al.\ 2003), the transfer function will not be strongly
affected; for example even if the disc is truncated at 1000\,R$_{\rm
Sch}$ (larger than proposed) then for a 10\,M$_{\odot}$ black hole
this corresponds to a delay in onset of only 0.1\,ls.

The extended response in the model comes from the far side of the
outer disc.  This is rather strong because the disc is flared and so
the outer parts are illuminated much more effectively than the disc
centre by a central point source (as assumed by this model).  High
inclinations and thick discs tend to reduce the visibility of the near
side of the disc, hence strengthening this extended response relative
to the peak.  This component is clearly not dominant in the
reconstructed transfer functions.  We have constructed simulated
lightcurves using the model transfer functions and then performed MEM
deconvolution.  The model is reconstructed very well, so if this were
the correct transfer function we should have recovered it.  Of course,
disc reprocessing might still contribute, together with the companion
star, to the extended response which does seem to be present up to
$\sim10$\,s, but something else is needed for the main, sharp peak.

\subsection{Disc transfer functions with a raised X-ray source}

We also considered a model in which the X-rays originate well above
the disc (a `lamppost'), possible if X-rays originate from
Comptonisation of disc EUV photons, or direct X-ray synchrotron, at
the base of a jet.  Both a $\sim1$\,s delay of the peak and the lack
of extended response could be resolved in this way.  The light travel
time from the Comptonisation region down to the disc will offset the
main peak.  By moving the X-ray source above the disc, the inner (UV
bright) region is more effectively illuminated and so this region can
contribute much more to the reprocessing than the outer disc, hence
suppressing the extended response.  This model suffers from the same
problem as the standard disc model, though; at the inferred high
inclination of this source, the first response to be seen is the long
wavelengths from the near-side of the disc.  This wavelength
dependence is the opposite of that observed.

\subsection{Disc emission lines?}
\label{LineEchoSection}

We have argued that the continuum-dominated transfer functions
discussed so far do not appear to be associated with thermal
reprocessing.  The UV emission lines could be, however; \target\ shows
strong emission from N\,{\sc v} 1240\,\AA\ and weaker lines of
Si\,{\sc iv} 1400\,\AA\ and He\,{\sc ii} 1640\,\AA\ (Haswell et al.\
2002).  Allowing for the doublet nature of N\,{\sc v} and Si\,{\sc
iv}, these lines all have a double peaked structure reminiscent of
disc lines (Haswell et al.\ in preparation).  Their most likely origin
is then in reprocessing the disc, so we would expect them to exhibit
lags of up to 7\,s as estimated above.  Since the main continuum
signal is concentrated at short lags, we might hope to separate
disc reprocessing at larger lags; we know from
Section~\ref{VarspecSection} that the N\,{\sc v} line is variable, and
other lines probably are.

\begin{figure}
\hspace*{3mm}\epsfig{angle=90,width=3.4in,file=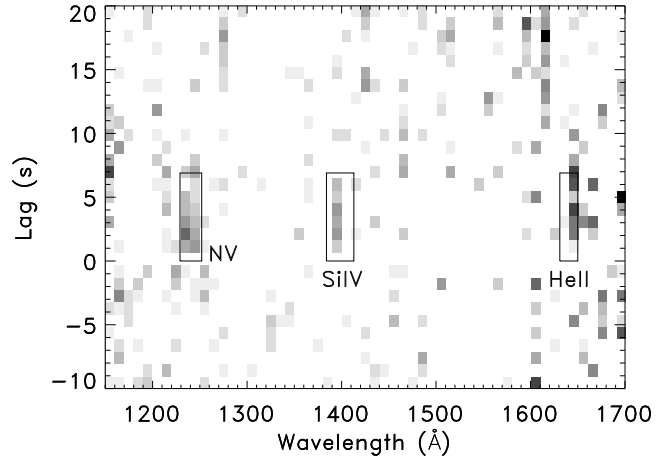}
\vspace*{0mm}
\caption{Wavelength dependent lags in the far-UV.  All are measured
with respect to the average lightcurve, and the average
cross-correlation function has been subtracted.  See text for details.
The boxes marked have widths corresponding to the measured FWZI of the
lines, and extend from 0--6.9\,s, corresponding to the expected range
of disc lags (with respect to the continuum).  Of these lines, N\,{\sc
v} is the strongest and has the most convincing detection, although
all are marginal.}
\label{MWCCFFig}
\end{figure}

To attempt to isolate this signal, we cross correlate wavelength
dependent lightcurves with the average lightcurve, as this should pick
out wavelengths with different variability properties to the average.
The result is a plot of the strength of the correlation as a function
of wavelength and lag.  Since the UV variability is very weak, this
proved extremely difficult and it was necessary to use rather crude
resolution (10\,\AA\ and 1\,s) and perform an exposure-time weighted
average of all these cross-correlations from all visits.  The result
is, as might be expected, dominated by the auto-correlation of the
continuum signal.  To remove this, we normalise each wavelength to the
same peak correlation and subtract the average.  The resulting
wavelength dependent residuals then reveal any wavelengths behaving
differently to the average.  Assuming that the lag of the far-UV
continuum (w.r.t.\ X-rays) is close to zero, and that the disc is
tidally truncated, we would expect continuum--line lags of up to 7\,s.
A larger continuum lag, or smaller disc, would both reduce this.  We
show the results of this analysis in Fig.~\ref{MWCCFFig}.  A response
appears at wavelengths corresponding to the strongest line, N\,{\sc
v}, and with the expected lag.  Features may be seen in the other
lines, but in all lines this is at best a marginal detections of
lagged variability.  Nonetheless, the data do appear consistent with
the far-UV lines originating in disc reprocessing, and the lags do
appear to be consistent with this interpretation given the assumed
system parameters.

%%%%%%%%%%%%%%%%%%%%%%%%%%%%%%%%%%%%%%%%%%%%%%%%%%%%%%%%%%%%%%%%%%%%%%%%%%%%%%%
%
\section{Discussion}
\label{DiscussionSection}
\subsection{Similarities to GX\,339--4}

The only other BHXRT that we know in which similar multi-wavelength
variablity has been adequately studied is GX\,339--4.  Fast optical
variations were seen in a low/hard X-ray state in 1981 May, when the
source was optically bright (Motch, Ilovaisky \& Chevalier 1982).  A
very pronounced QPO was present around 0.05\,Hz, with a full amplitude
of 30--40\,percent.  The frequency appeared stable over the two nights
observed.  Very fast flares were also present on timescales of
$\sim10$\,ms with amplitudes of up to a factor of 5.  Overall the
power spectrum looks strikingly similar to those we see, with
band-limited noise and a strong QPO at slightly above the break
frequency.  Contemporaneous X-ray timing observations were also made,
including a 96\,s segment simultaneous with optical coverage (Motch et
al.\ 1983).  X-ray and optical power-spectra have similar structure
but subtle differences.  The 0.05\,Hz QPO is present, but there is an
equally strong harmonic at 0.1\,Hz.  The high frequency power-spectrum
also drops off somewhat faster in the X-rays (i.e.\ the slope is
steeper, as is the case for \target; see Table~\ref{HSTPSDTable}).  No
sharp positive correlation is seen in the X-ray--optical CCF, but a
broad dip is present.  The presence of a strong QPO introduces some
ambiguity in interpreting the CCF, but Motch et al.\ (1983) suggest
the most likely interpretation to be that the optical is
anticorrelated with the X-rays and leads by $2.8\pm1.6$\,s.
Examination of the simultaneous lightcurves does support this,
suggesting the presence of optical dips preceding with X-ray flares,
although the match-up is not perfect since some optical dips are not
associated with X-ray flares.  These dips may be analogues of the
precognition dips in \target.  The short optical flares, however, do
not seem to be correlated with X-ray events, just as in \target\ we
find that the X-ray/UV coherence decreases at frequencies $\ga0.1$\,Hz
(Fig.~\ref{CrossSpecFig}).

The optical QPO was seen again with a frequency in the range
0.05--0.20\,Hz in 1982 May (Motch et al.\ 1985), 1989 August (Imamura
et al.\ 1990), and 1996 April (Steiman-Cameron et al.\ 1997).  In all
of these cases the QPO was weaker, with amplitude $<10$\,percent, and
the source was optically fainter.

For the fastest flares, Motch et al.\ (1982) derived a
brightness temperature of $5\times 10^{9}$\,K and suggested an origin
either in bremsstrahlung or in cyclotron or synchrotron emission.
Fabian et al.\ (1982) argued that the X-ray brightness was too low for
the bremsstrahlung interpretation, that the optical flux originates
in cyclotron emission, which is optically thick into the optical band,
and that the QPO could come from variations in the size of the inner
hot region responsible for the emission.  They estimated an infall
time for the inner region of order a few seconds and suggested that
Compton cooling of hot regions by soft X-rays could result in the
optical dips.  Apparao (1984), however, argued that this model does
cannot reproduce the inferred luminosity of the optical cyclotron
emission, and instead suggested that the optical emission arises from
reprocessing of {\em hard} X-rays by the disc.  This is certainly not
a complete review of models for low/hard state variability proposed
for GX\,339--4, but indicates the main ingredients that have been
proposed.  

We have summarised these results and conclusions because there are
some striking similarities to \target.  The strong optical
variability, similarity in form of the X-ray and optical
power-spectra, and presence of optical QPOs are all characteristic of
both sources.  The QPO frequency ranges spanned are similar, and the
X-ray--optical anticorrelation reported by Motch et al.\ (1983) is
reminiscent of the `precognition dip' seen in \target.  Observations
of both sources occurred in X-ray low/hard states, which exhibited a
relatively high optical luminosity.  It is then natural to look for a
common interpretation of the phenomena seen in both sources, and as
for GX\,339--4, we may have to look to cyclotron or synchrotron
emission to explain the UVOIR variability in \target.

\subsection{The origin of the UV/optical/IR variability}

The properties of the UVOIR variability do not appear consistent with
it being dominated by thermal reprocessing either in the disc or on
the companion star.  Most obviously, the variability amplitude
increases at longer wavelengths where we expect the disc contribution
to be minimised, but the correlation properties are also difficult to
interpret in this way.  The main signal does not vary with phase as
expected for the companion star, and has a shape, and wavelength
dependence, which is inconsistent with models of reprocessing in a
disc.  These conclusions, that something other than reprocessing
dominates the UVOIR variability, are the same as those of Kanbach et
al.\ (2001), but based on different arguments.  They argued that the
narrowness of the optical ACF proved reprocessing is not responsible.
This is not necessarily true, however; provided a negative response is
included sharp optical ACFs are in principal possible, {\em even if
the main peak were due to reprocessing}.  We have presented a more
rigorous test and falsification of the reprocessing hypothesis.

Instead several different mechanisms appear to be responsible for
UVOIR variability.  Reprocessing may produce a weak continuum signal
moving with the companion star, and possibly a weak signal from the
disc as well.  Disc reprocessing probably is detected in the UV
emission lines.  None of these components dominate the correlated
variability.  Another mechanism is needed to explain the main peak,
and the precognition dip.  These effects might be closely associated
or relatively independent.  Spruit \& Kanbach (2002) have performed a
principal component analysis of time-resolved CCFs and find that the
dip and the peak are statistically correlated, and appear to be two
aspects of the same process.  In GX\,339--4, however, a similar
precognition dip is seen, but there is no strong positive correlation.
Some evidence for identifying the main variability mechanism as
synchrotron comes from the spectral energy distribution of the
variability, which exhibits a single power-law form consistent with
optically thin synchrotron emission, and extending to the X-ray regime
which is usually interpreted as due to inverse-Compton emission.

Any model for the main peak of the response must explain why it is
progressively more lagged at longer wavelengths.  This is the case
whether the lags are measured directly from the CCF or from a transfer
function reconstructed in several ways, and so is a robust result.  It
is insensitive to the accuracy of the absolute timing, since the
systematic difference between wavelengths is greater than the scatter
at a single wavelength.  This effect does not straightforwardly emerge
from models in which the correlation is due to scattering, but would
be expected if the UVOIR variability comes from expanding synchrotron
bubbles, which become optically thin at progressively longer
wavelengths with time.  As we have already argued, this wavelength
dependence is not expected in a reprocessing model, as due to the high
inclination, the first reprocessed signal to be received should
actually be the long wavelengths from the near-side of the outer disc.

A model for the precognition dip must explain why it seems stronger
for the longer wavelength data; this trend is present within our data,
and is particularly  clear when compared to optical results of Kanbach et
al.\ (2001) who find a very pronounced dip on the optical.  The dips
were similarly pronounced in the optical observations of GX\,339--4.
One approach is to argue that there are two components to the
synchrotron emission, a persistent level and rapid flares.  An optical
dip preceding an X-ray flare might then represent a reduction in the
persistent synchrotron emission.  Fabian et al.\ (1982) suggested that
this could be due to Compton cooling by soft X-ray photons, removing
the high energy electrons.  But this should produce a more pronounced
dip at short wavelengths which arise from more energetic electrons.
Spruit \& Kanbach (2002) suggested an alternative in the context of a
reprocessing model, that preceding an X-ray flare, a region of the
inner disc puffs up somewhat and shields the outer disc from
irradiation.  In this case too, however, the dip should be associated
with the blue disc component.  An alternative explanation is that the
dips are due to increasing self-absorption (Kaiser, 2002, priv.\
comm.).  This could occur if the electron density increased, for
example because a cloud of plasma is shrinking.  Since the
self-absorption will be stronger at longer wavelengths, we would
expect the dips to be more pronounced there as observed.  The
timescales of the dips, a few seconds, correspond to a few times the
dynamical timescale of an inner evaporated region extending to
$\sim350$\,R$_{\rm Sch}$ (Chaty et al.\ 2003), so this is
not implausible.

Another requirement for a model is that it must include a mechanism to
produce low-frequency QPOs.  These are ubiquitous in both \target\ and
GX\,339--4, and observed in X-rays, optical, and UV.  It is likely
that they are associated with the synchrotron component, not just with
disc emission, but they could either represent changes in the
intrinsic emission, or variable obscuration of the emitting region by
a structure in the disc.  The latter explanation seems unlikely, as
QPOs generally become stronger at higher energies, whereas an
obscuration model would predict more absorption at lower energies.
The frequency is not stable on long timescales but does generally
evolve smoothly and relatively slowly; it is not subject to large
fluctuations.  It does not vary monotonically with X-ray luminosity
(Wood et al.\ 2000).  One `clock' which might vary in this way would
be the inner disc radius (c.f.\ Merloni et al.\ 2000).  If the QPO
frequency varied proportionally to the Keplerian frequency at the
inner disc edge, or to the size of an inner advective region, then a
decrease in the inner radius could produce the observed evolution.
Alternatively, Wood et al.\ (2001) have suggested that the QPO
frequency is inversely related to the total disc mass.

A final observational characteristic relates to the coherence of X-ray
and optical/UV variations.  It is clear from the broad-band PSD that a
range of flare timescales is involved in the variability, although the
QPO may define a preferred timescale.  As one moves to higher
frequencies of variability, in both \target\ and GX\,339--4 the
coherence decreases, and whereas low frequency variations involve
correlated X-ray and optical/UV behaviour, at high frequencies only
the optical/UV participates, or at least the flares become
uncorrelated.  This may indicate a difference in the emission
properties of the shortest flares (presumably associated with smaller
flaring regions) relative to longer ones.

None of these comments are intended to define a detailed model but
only to indicate the key observational facts that a model must
explain, and to suggest ingredients which might contribute to such a
model.  A number of spectral models have been advanced for this
source, involving a magnetically dominated corona above the disc
(Merloni et al.\ 2000); jet emission (Markoff et al.\ 2001); an
advection dominated accretion flow (ADAF; Esin et al.\ 2001); and an
inner, hot disc (Frontera et al.\ 2001).  However, only Merloni et
al.\ (2000) explicitely address the variability properties expected
and their model has a number of difficulties, both with the predicted
spectra and the variability properties.  It assumes an accretion disc
extending to the last stable orbit and hence predicts a large hump in
the soft X-ray regime; this is not consistent with the \EUVE\ and
\Chandra\ data (Hynes et al.\ 2000; McClintock et al.\ 2001b).  The
authors note, however, that there are similarities to an ADAF model,
and discuss the expected variability in both cases.  As they predict,
optical variability does extend to timescales of tens of milliseconds
(as is also the case in GX\,339--4), but whereas they expect the
variability to drop off strongly at long wavelengths, we actually see
the opposite, and even in the infrared large amplitude variations are
present.  This is arises in their model because the cyclo-synchrotron
emission is fully self-absorbed.  The spectral data (Chaty et al.\
2003) and the extension of variability to long wavelengths, however,
suggest that the synchrotron emission has a flatter spectrum likely
indicating an inhomogeneous medium with local spectra peaking at a
wide range of self-absorption frequencies.  A final difficulty of this
model is that the lags have the wrong sense if X-rays are produced by
Comptonisation of the synchrotron emission.  It is also worth
remarking on the jet model of Markoff et al.\ (2001).  To fit the flat
UVOIR spectrum, these authors use a combination of optically thin
synchrotron, becoming optically thick in the IR, and disk emission.
If this model is correct, then our broad-band variability spectrum
appears to have very effectively isolated the synchrotron component.
%
%%%%%%%%%%%%%%%%%%%%%%%%%%%%%%%%%%%%%%%%%%%%%%%%%%%%%%%%%%%%%%%%%%%%%%%%%%%%%%%
%
\section{Conclusions}
We identify short timescale variability ($\la100$\,s) at all energies
where it would be detectable, in X-rays, UV, optical, and infrared.
The variability amplitude is very high at X-ray and infrared energies,
moderate in the optical and lowest in the UV.  This suggests that the
UVOIR variability is associated with the synchrotron component usually
invoked to explain the long wavelength spectrum of this and similar
sources, rather than with the disc which is expected to have a blue
spectrum. Indeed, the broad-band variability spectrum is
consistent with optically thin synchrotron emission.

To explore this we have analysed and compared the variability in
different bands.  All of our data exhibit power density spectra
consistent with band-limited noise typical of sources in the low/hard
state, although the time resolution of the IR data is too low to
detect the break in the PSD.  A QPO is detected in some X-ray, UV, and
optical data.  The frequency and other properties of the QPO are
consistent across all bands and the QPO evolves
monotonically to higher frequencies during the outburst as previously
found from X-ray data alone.

The X-ray and UV data, for which we have high time-resolutions and
several periods of overlap, show clear correlations.  The UV lags
behind the X-rays, with slightly larger lags at longer wavelengths.
The properties of the correlation are puzzling, however.  As
previously noted by Kanbach et al.\ (2001), the UV/optical ACFs are
narrower than the X-ray ones, and the CCF rises very rapidly.  This is
not expected from disc reprocessing, for which positive, linear
responses are expected and hence the optical/UV ACF should be
broadened with respect to the X-ray one.  It can however arise if
there is at times a negative response.  Transfer functions including a
`precognition dip' (Kanbach et al.\ 2001) can reproduce these
properties.  Even when this negative component is minimised, the main
peaks of the transfer functions are inconsistent with disc
reprocessing, and do not move from observation to observation as
expected for reprocessing on the companion star.  Combined, these
properties of the optical/UV continuum variability are inconsistent
with an origin for the main signal in thermal reprocessing anywhere in
the system, and support a model involving synchrotron variability, or
some other non-thermal optical/UV source.  We know of no detailed
model which can fully account for the observed multi-wavelength
properties of the variability.

Continuum reprocessing from the disc may be present as part of the
weak extended response, and a weak feature moving with the companion
star may also be present.  UV spectra indicate that both line and
continuum are variable.  There is marginal evidence that the lines may
be delayed with respect to the continuum by up to 10\,s.  This is
consistent with reprocessing in the disc, and indeed the lines do have
a double-peaked disc profile.

To summarise, the UVOIR variability in \target\ is complex, with
several components.  We suggest that the most prominent components, a
dip and peak leading and lagging the X-rays respectively are
associated with synchrotron variability.  Thermal reprocessing may
also be present and is suggested by the UV emission lines and a weak
component of the continuum variability apparently moving with the
companion star.
%
%%%%%%%%%%%%%%%%%%%%%%%%%%%%%%%%%%%%%%%%%%%%%%%%%%%%%%%%%%%%%%%%%%%%%%%%%%%%%%%
%
\section*{Acknowledgements}
RIH would like to thank Ross Collins, Heino Falcke, Christian Kaiser,
Julien Malzac, Sera Markoff, Danny Steeghs, and Phil Uttley for useful
scientific discussions and an anonymous referee for careful reading
and good ideas.  Also thanks to Reba Bandyopadhyay, on behalf of the
USA team, for providing the {\it ARGOS}/USA data reproduced in
Fig.~\ref{LCFig}.

This work includes observations with the NASA/ESA {\it Hubble Space
Telescope}, obtained at the Space Telescope Science Institute, which
is operated by the Association of Universities for Research in
Astronomy, Inc.\ under NASA contract No.\ NAS5-26555.  We would like
to thank all at STScI for their continued support, especially Tony
Roman for implementation of the observations and Kailash Sahu, Charles
Proffitt, and David Stys for patient assistance with technical
difficulties.  RIH would also like to thank Ted Gull for an
informative discussion on the accuracy of \HST\ timing.  We would also
like to thank the \RXTE\ team for their support and especially the
schedulers for facilitating simultaneous observations.

The United Kingdom Infrared Telescope is operated by the Joint
Astronomy Centre on behalf of the UK Particle Physics and Astronomy
Research Council.  SC would like to thank John K. Davies, Andy Adamson
and Sandy K. Leggett who scheduled all the override observations and
kindly gave advice for their set-up. SC would like especially to
acknowledge John K. Davies and Sandy K. Leggett who took the
observations of June 24 and July 15 respectively.  SC is also very
grateful to J. Mart\'{\i} who triggered the June 24 observations on
his behalf.

RIH, CAH and SC acknowledge support from grant F/00-180/A from the
Leverhulme Trust.  In addition RIH acknowledges financial aid from
NOVA for part of this work.  RIH is currently supported by NASA
through Hubble Fellowship grant \#HF-01150.01-A awarded by the Space
Telescope Science Institute, which is operated by the Association of
Universities for Research in Astronomy, Inc., for NASA, under contract
NAS 5-26555.  WC would like to acknowledge NASA LTSA grant NAG5-7990.
Support for \HST\ proposal GO\,8647 was provided by NASA through a
grant from the Space Telescope Science Institute.

This work has made use of the NASA Astrophysics Data System Abstract
Service.

\label{lastpage}


\begin{thebibliography}{99}
%
\bibitem{Apparao:1984a}
  Apparao K. M. V., 1984, A\&A, 139, 375
%
\bibitem{Bendat:1986a}
  Bendat J., Piersol A., 1986, Random Data: Analysis and
  Measurement Procedures, Wiley, New York
%
\bibitem{Blandford:1982a}
  Blandford R. D., McKee C. F., 1982, ApJ, 255, 419
%
\bibitem{Chaty:2003a}
  Chaty S., Haswell C. A., Malzac J., Hynes R. I, Shrader C. R., Cui W., 
  2003, MNRAS, submitted
%
\bibitem{Cherepashchuk:2000a}
  Cherepashchuk A. M., 2000, Space Science Reviews, 93, 473
%
\bibitem[Cook et al.(2000)]{C2000}
  Cook L., Patterson J., Buczynski D., Fried R., 2000, 
  IAU Circ.\ 7397
%
\bibitem{Edelson:1988a}
  Edelson R. A., Krolik J. H., 1988, ApJ, 333, 646
%
\bibitem{Esin:2001a}
  Esin A. A., McClintock J. E., Drake J. J., Garcia M. R.,
  Haswell C. A., Hynes R. I., Muno M. P., 2001, ApJ, 555, 483
%
\bibitem{Fabian:1982a}
  Fabian A. C., Guilbert P. W., Motch C., Ricketts M., Ilovaisky
  S. A., Chevalier C., 1982, A\&A, 111, L9
%
\bibitem{Fender:2001a} 
  Fender R. P., Hjellming R. M., Tilanus R. P. J., 
  Pooley G. G., Deane J. R., Ogley R. N., Spencer R. E., 
  2001, MNRAS, 322, L23 
%
\bibitem{Ferguson:1999a}
  Ferguson H., Baum S., 1999, Scientific Requirements for Thermal
  Control and Scheduling of the STIS MAMA detectors after SM-3,
  STIS ISR 99-02, STScI
%
\bibitem{Frontera:2001a}
% 15 authors abbreviated as Frontera et al.
Frontera F., et al., 2001, ApJ, 561, 1006 
%
\bibitem[Garcia et al.(2000)]{G2000}
  Garcia M., Brown W., Pahre M., McClintock J., Callanan
  P., Garnavich P., 2000, IAU Circ.\ 7392
%
\bibitem{Gull:1998a}
  Gull T. R., et al., 1998, ApJ, 495, L51
%
\bibitem{Haswell:2000a}
  Haswell C. A., Skillman D., Patterson J., Hynes R. I., Cui W.,
  2000, IAU Circ.\ 7427
%
\bibitem{Haswell:2002a}
  Haswell C. A., Hynes R. I., King A. R., Schenker K., 2002, 
  MNRAS, 332, 928
%
\bibitem{Hawarden:2001a}
  Hawarden T. G., Leggett S. K., Letawsky M. B., Ballantyne
  D. R., Casali M. M., 2001, MNRAS, 325, 563
%
\bibitem{Horne:1994a}
  Horne K., 1994, in Gondhalekar P. M., Horne K., Peterson
  B.M., eds., Reverberation Mapping of the Broad-Line Region in
  Active Galactic Nuclei. ASP Conf.\ Series, Vol.\ 69, p23
%
\bibitem{Horne:1991a}
  Horne K., Welsh W. F., Peterson B. M., 1991, ApJ, 367, L5
%
\bibitem{Hynes:1998a}
  Hynes R. I., O'Brien K., Horne K., Chen W., Haswell C. A., 
  1998, MNRAS, 299, L37
%
\bibitem{Hynes:2000a} 
  Hynes R. I., Mauche C. W., Haswell C. A., Shrader C. R., Cui
  W., Chaty S., 2000, ApJ, 359, L37
%
\bibitem[\protect\citename{Ilovaisky et al.\ }1980]{I80}
  Ilovaisky S. A., Chevalier C., White N. E., Mason K. O.
  Sanford P. W., Delvaille J. P., Schnopper H. W., 1980, 
  MNRAS, 191, 81
%
\bibitem{Imamura:1990a}
  Imamura J. N., Kristian J., Middleditch J., Steiman-Cameron
  T. Y., 1990, ApJ, 365, 312 
%
\bibitem{Kanbach:2001a} 
  Kanbach G., Straubmeier C., Spruit H. C., Belloni T., 2001, 
  Nat, 414, 180
%
\bibitem{Kazanas:1999a}
  Kazanas D., Hua, X. -M., 1999, ApJ, 519, 750
%
\bibitem{Leahy:1983a} 
Leahy D. A., Darbro W., Elsner R. F., Weisskopf M. C., Kahn S., 
Sutherland P. G., Grindlay J. E., 1983, ApJ, 266, 160 
%
\bibitem{Leitherer:2001a}
% Full author list not given, likely large.  Referred to as Leitherer
% et al. at specific request of STScI
  Leitherer, C., et al., 2001, STIS Instrument Handbook, Version
  5.0, STScI, Baltimore
%
\bibitem[Lockman, Jahoda \& McCammon(1986)]{LJM1986}
  Lockman F. J., Jahoda K., McCammon D., 1986, ApJ, 302, 432
%
\bibitem{Long:2000a}
  Long C., 2000, STIS Time-Tag Timing, STIS Technical White
  Paper 00-175, STScI, Baltimore
%
\bibitem{McClintock:2001a} 
  McClintock J. E., Garcia M. R., Caldwell N., Falco E. E.,
  Garnavich P. M., Zhao P., 2001a, 551, L147
%
\bibitem{McClintock:2001b}
% 16 authors abbreviated to McClintock et al.
  McClintock J. E. et al., 2001b, ApJ, 555, 477
%
\bibitem{Markoff:2001a}
  Markoff S., Falcke H., Fender R., 2001, A\&A, 372, L25 
%
\bibitem{Merloni:2000a}
  Merloni A., Di Matteo T., Fabian A. C., 2000, MNRAS, 318, L15
%
\bibitem[\protect\citename{Middleditch \& Nelson }1976]{MN76}
  Middleditch J., Nelson J. E., 1976, ApJ, 208, 567
%
\bibitem{Miller:2002a} 
Miller J. M., Ballantyne D. R., Fabian A. C., Lewin W. H. G., 2002, 
MNRAS, 335, 865 
%
\bibitem{Motch:1982a}
  Motch C., Ilovaisky S. A., Chevalier C., 1982, A\&A, 109, L1
%
\bibitem{Motch:1983a}
  Motch C., Ricketts M. J., Page C. G., Ilovaisky S. A.,
  Chevalier C., 1983, A\&A, 119, 171
%
\bibitem{Motch:1985a}
  Motch C., Ilovaisky S. A., Chevalier C., Angebault P., 1985,
  SSR, 40, 219
%
\bibitem{OBrien:2001a} 
  O'Brien K., Horne K., 2001, in proc.\ Astro-Tomography, eds.\ 
  H. Boffin, D. Steeghs, J. Cuypers, Springer-Verlag Lecture 
  Notes in Physics Series, p416
%
\bibitem{OBrien:2002b}
  O'Brien K., Horne K., Hynes R. I., Chen W., Haswell C. A.,
  Still M. D., 2002, MNRAS, 334, 426
%
\bibitem{Orosz:2001a} 
  Orosz J. A., 2001, ATEL \#67
%
\bibitem[\protect\citename{Petro et al.\ }1981]{P81}
  Petro L. D., Bradt H. V., Kelley R. L., Horne K., Gomer R.,
  1981, ApJ, 251, L7
%
\bibitem[Pooley \& Waldram(2000)]{PW2000}
  Pooley G. G., Waldram E. M., 2000, IAU Circ.\ 7390
%
\bibitem[Remillard et al.(2000)]{R2000}
  Remillard R., Morgan E., Smith D., Smith E., 2000, 
  IAU Circ.\ 7389
%
\bibitem{Revnivtsev:2000a}
  Revnivtsev M., Sunyaev R., Borozdin K., 2000, A\&A, 361, L37
%
\bibitem{Skillman:1993a}
  Skillman D. R., Patterson J., 1993, ApJ, 417, 298 
%
\bibitem{Spruit:2002a}
  Spruit H. C., Kanbach G., 2002, A\&A, 391, 225
%
\bibitem{SteimanCameron:1997a}
  Steiman-Cameron T. Y., Scargle J. D., Imamura J. N.,
  Middleditch J., 1997, ApJ, 487, 396
%
\bibitem[Tanaka \& Shibazaki(1996)]{TS1996}
  Tanaka, Y., Shibazaki, N., 1996, ARA\&A, 34, 607
%
\bibitem[Uemura, Kato \& Yamaoka(2000)]{UKY2000}
  Uemura M., Kato T., Yamaoka H., 2000, IAU Circ.\ 7390
%
\bibitem[Uemura et al.(2000)]{U2000} 
% 10 authors abbreviated to Uemura et al.
  Uemura M. et al., 2000, PASJ, 52, L15
%
\bibitem{Vaughan:1997a}
  Vaughan B. A., Nowak M. A., 1997, ApJ, 474, L43
%
\bibitem{vanderKlis:1995a}
  van der Klis M., 1995, in X-ray Binaries, eds.\
  W. H. G. Lewin, J. van Paradijs, E. P. J. van den Heuvel, CUP,
  p.\ 252
%
\bibitem{vanderKlis:2000a} 
  van der Klis M., 2000, ARA\&A, 38, 717
%
\bibitem{Wagner:2001a} 
  Wagner R. M., Foltz C. B., Shahbaz T., Casares J., Charles P. A., 
  Starrfield S. G., Hewett P., 2001, ApJ, 556, 42
%
\bibitem{Wijnands:1999a}
  Wijnands R., van der Klis M., 1999, ApJ, 514, 939
%
\bibitem[Wilson \& McCollough(2000)]{WM2000}
  Wilson C. A. McCollough M. L., 2000, IAU Circ.\ 7390
%
\bibitem{Wood:2000a}
% 16 authors abbreviated to Wood et al.
  Wood K. S., et al., 2000, ApJ, 544, L45
%
\bibitem{Wood:2001a}
  Wood K. S., Titarchuk L., Ray P. S., Wolff M. T., 
  Lovellette  M. N., Bandyopadhyay R. M., 2001, ApJ, 563, 246 
%
\bibitem{Zurita:2002a}
% 11 authors abbreviated to Zurita et al.
  Zurita C., et al., 2002, MNRAS, 333, 791
%
\end{thebibliography}
\end{document}